%% file: main.tex
\documentclass[9pt,conference]{IEEEtran}
\usepackage{graphicx,float}
\usepackage{caption}
\usepackage[square, numbers]{natbib}
\usepackage{hyperref}
\usepackage{listings}
\usepackage{booktabs}
\usepackage{amsmath} 
\usepackage{array}
\usepackage{float}

\hypersetup{colorlinks,linkcolor=blue,filecolor=blue,urlcolor=blue,citecolor=blue}
\def\docroot{.}
\graphicspath{{\docroot/Figures/}}

\newcommand{\tabhead}[1]{\textbf{#1}}

\newcommand{\PreserveBackslash}[1]{\let\temp=\\#1\let\\=\temp}
\newcolumntype{C}[1]{>{\PreserveBackslash\centering}p{#1}}
\newcolumntype{R}[1]{>{\PreserveBackslash\raggedleft}p{#1}}
\newcolumntype{L}[1]{>{\PreserveBackslash\raggedright}p{#1}}
\title{Evaluation of NoSQL in the Energy Marketplace with GraphQL Optimization}

\author{
    \IEEEauthorblockN{Michael Howard P.Eng}
    \IEEEauthorblockA{\textit{Department of Computer Science} \\
        \textit{Maseeh College of Engineering and Computer Science }\\
        Portland State University \\
        email: mihoward@pdx.edu
    }
}
\begin{document}

\maketitle
\thispagestyle{plain}
\input{Abstract/Abstract}
\input{Chapters/Chapter1}
\input{Chapters/Chapter2}
\input{Chapters/Chapter3}
\input{Chapters/Chapter4}

\input{Chapters/Chapter5}
\input{Chapters/Chapter6}

\newpage
\section*{Acknowledgements}
The author would like to thank Dr. Wu-chang Feng, Dr. R. Bruce Irvin and Dr. David Maier at Portland State University for their assistance, valuable advice, direction and mentorship provided throughout this study.

\bibliography{References/References}

\end{document}

%% file: Abstract/Abstract.tex
\begin{abstract}
The growing popularity of electric vehicles in the United States requires an ever-expanding infrastructure of commercial DC fast charging stations.  The U.S. Department of Energy estimates 33,355 publicly available DC fast charging stations as of September 2023.  In 2017, 115,370 gasoline stations were operating in the United States, much more ubiquitous than DC fast chargers.  Range anxiety is an important impediment to the adoption of electric vehicles and is even more relevant in underserved regions in the country.  The peer-to-peer energy marketplace helps fill the demand by allowing private home and small business owners to rent their 240 Volt, level-2 charging facilities.  The existing, publicly accessible outlets are wrapped with a Cloud-connected microcontroller managing security and charging sessions.  These microcontrollers act as Edge devices communicating with a Cloud message broker, while both buyer and seller users interact with the framework via a web-based user interface.  The database storage used by the marketplace framework is a key component in both the cost of development and the performance that contributes to the user experience.  A traditional storage solution is the SQL database.  The architecture and query language have been in existence since the 1970s and are well understood and documented.  The Structured Query Language supported by the query engine provides fine granularity with user query conditions.  However, difficulty in scaling across multiple nodes and cost of its server-based compute have resulted in a trend in the last 20 years towards other NoSQL, serverless approaches.  In this study, we evaluate the NoSQL vs. SQL solutions through a comparison of Google Cloud Firestore and Cloud SQL MySQL offerings.  The comparison pits Google's serverless, document-model, non-relational, NoSQL against the server-base, table-model, relational, SQL service.  The evaluation is based on query latency, flexibility/scalability, and cost criteria.  Through benchmarking and analysis of the architecture, we determine whether Firestore can support the energy marketplace storage needs and if the introduction of a GraphQL middleware layer can overcome its deficiencies. 

\end{abstract}

\begin{IEEEkeywords}
    Non-relational, relational, MySQL, mitigate, Firestore, SQL, NoSQL, serverless, database, GraphQL
\end{IEEEkeywords}

%% file: Chapters/Chapter1.tex
\section{Introduction}\label{sec:introduction}
The era of internal combustion vehicles is coming to a close.  Electric vehicle (EV) sales are on a steady upward trajectory. EV-Volumes \cite{Irle2023} report that in 2022, sales increased by 55\% over the previous year, while the entire auto industry declined by -0.5\% during that same period.  A major barrier to the ownership of electric vehicles is range anxiety.  Commercial charging facilities are not yet as ubiquitous as gasoline stations, opening up a market for private individuals to assist by sharing their personal charging facilities.  In this paper, we detail how the Cloud framework for a peer-to-peer energy marketplace might work to provide users with a means to share and rent their private charging outlets.  Components of the framework include the Edge microcontrollers, web server, API, Cloud Functions, Publication-Subscription service, and a message broker.  We then dig deep into the database component, which is the main research question. 

\subsection{Motivations}\label{sub:motivations}
\begin{figure}[H]
    \centering
    \includegraphics[width=.9\linewidth, keepaspectratio]{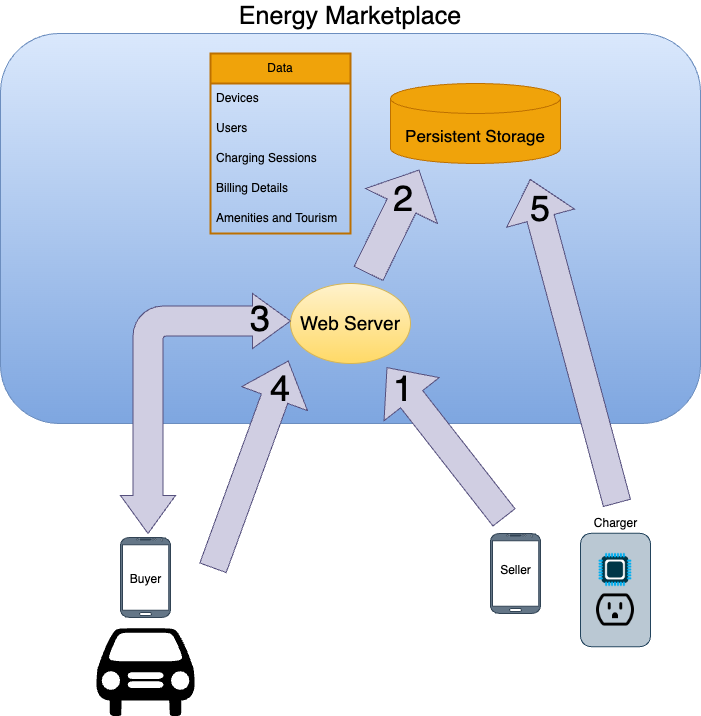}
    \captionsetup{width=.8\linewidth}
    % \decoRule
    \caption{The energy marketplace supporting: 1. Device registration, 2. Data storage, 3. Device search and reserve, 4. Session management, and 5. Session details.}
    \label{fig:energy-marketplace-workflow}    
\end{figure}

A critical step in the energy marketplace is to persistently store data objects that represent the Edge charging devices. Users, charging session history, billing details, ratings and reviews, amenities, and tourist information are additionally persisted.  The workflows for the marketplace users include registering a new charger device, searching for available devices, and managing a charging session, as seen in Figure \ref{fig:energy-marketplace-workflow}.  

Requirements for persistent data include low read and write latency, the ability to scale horizontally as data grows, the cost, and the ease of evolving the shape of the data.   Multiple users may query persistent storage, asking for data objects that contain multiple fields within a given range.  Read-operation latency will be critical in this scenario.  The user must be able to request a charger within a given latitude and longitude radius and may need to include the maximum distance to a restroom, the minimum user rating, the type of charger, etc.  The query engine must be able to process these patterns of field conditionals.  If it can process only a subset, the resulting data reduction must be efficient enough so that the user response is not noticeably delayed.

Another motivation for investigating database technology is flexibility and scalability.  The energy marketplace will grow iteratively, adding features that require additional persistence.  A fixed schema or data shape may require a migration (or a new table) of all existing persisted data to match the updated schema prior to adding the new data objects.  A flexible schema may reduce the burden on administrators and reduce administration complexity, since new fields can be incrementally stored as the marketplace framework and feature set evolve.  Since the number of users participating in the marketplace is also evolving, persistent storage must be able to horizontally scale to accommodate them.  For example, an early marketplace pilot may have 50 users within a single geographical zone.  An individual Cloud node with failover and redundancy can safely store all required persistent data and support all incoming query requests within the zone.  However, as users spread out geographically, storage nodes will need to scale into these new zones.  Also, as data and user request volume exceed what can be processed by a single node, storage must partition and scale across a node cluster. 

The high-level goals for the persistent storage for the marketplace are as follows:
\begin{enumerate}
    \item Storage cost.
    \item Compute cost.
    \item Query support.
    \item Query latency.
    \item Ease of scaling.
    \item Ease of data-structure evolution.
\end{enumerate}

Cost is a significant motivation, driving a design goal of the framework to avoid perpetual, unmanaged compute.  The deployment of server-based compute services may increase billing, as the compute resources are dedicated and always available.  Additionally, the type of compute and operating system patches must be managed by the developer.  A serverless persistent storage model provides a pricing model in which the billing only reflects what is used.  Thus, queries and storage contribute directly to billing, but idle time does not.  

\subsection{Research Question}\label{sub:question}
The aforementioned requirements for persistent storage raise the question of what will work best to support the energy marketplace.  A deep dive investigation is beneficial prior to developing the marketplace framework and leads us to the research question addressed in this paper: \textbf{Will a serverless, NoSQL, non-relational, document-model database (i.e. Google Firestore \cite{Firestore}) support the query, cost, and flexibility needs for persistent storage in a peer-to-peer energy marketplace?}  We will endeavor to answer this question through benchmarking latency, estimating costs, and analyzing the architecture.  To provide a comparison with a well-established solution, benchmarking and analysis are compared side-by-side against a relational, SQL, and perpetual server provided through Google Cloud SQL \cite{GoogleMySQL}.  

\subsection{Context of the Study}\label{sub:context}
Multiple related studies utilize a third-party benchmarking tool such as Yahoo! Cloud Serving Benchmark (YCSB) \cite{YCSB} to establish latency under variable load \cite{Cattell2011,Khazaei2016,Pandey2020,Kesavan2023}.  In this study, we will focus on a more customized way to measure the response time of the database.  

For both the relational and non-relational databases, we develop client applications that emulate portions of the energy marketplace data flow.  Queries running from these clients perform read-and-write operations that characterize latency performance in specific load scenarios.  This experimental setup allows the persistent storage component of the marketplace to be developed and tested in isolation.  
%The solution representing the best trade-offs around cost, latency and flexibility will continue to live on in the marketplace framework and support the investigation of future research problems. 

\subsection{Objectives and Contributions}\label{sub:objectives}
The objectives of this paper are as follows:
\begin{enumerate}
    \item Characterize query latency for both relational and non-relational databases.
    \item Evaluate the cost of both solutions.
    \item Explore the ease of evolving the stored data structure.
    \item Explore the ease of horizontal scaling.
    \item Evaluate inserting a GraphQL middleware layer and its effect on client simplicity, latency, safety, and portability.
\end{enumerate}

Achieving these objectives provides a characterization of the strengths and weaknesses of Firestore's persistent storage solution, both standalone and comparatively against the more established SQL solution.  This insight into Firestore's performance and capabilities contributes a known data storage platform to support the future energy marketplace framework.  

\subsection{Overview of the Paper}\label{sub:overview}
After initially introducing the topic, the background of electric vehicle (EV) charging and related research are covered in Section \ref{sec:background}.  Next, we discuss the system architecture of the Firestore and MySQL storage solutions, along with GraphQL, in Section \ref{sec:architecture}.  Section \ref{sec:experiments} presents the experiments performed against both solutions along with their corresponding experimental data.  The discussion of the experimental results, the regression model and future work is detailed in Section \ref{sec:discussion-future}.  Finally, we wrap up the paper with the conclusion in Section \ref{sec:conclusion}.

%% file: Chapters/Chapter2.tex
\section{Background}\label{sec:background}

\subsection{EV Charging}\label{sub:ev-charging}
Electric vehicle (EV) ownership has increased significantly in the United States.  At the end of 2022, there were 2,442,300 EV registrations compared to a combined 248,529,800 diesel and gasoline vehicles \cite{Doeregistrations2022}.  Thus, electric vehicles now represent approximately 1\% of all vehicles on the road. 

Accelerating electric vehicle sales requires an acceleration in the deployment of the charging infrastructure.  What if, by 2030, there was a peer-to-peer marketplace such that consumers could turn into producers and sell units of energy?  The growing popularity of solar panels \cite{Solar2022} allows the generation of energy and the reselling of energy purchased through a utility grid.  Peer-to-peer sales will help meet the growing demand for electric vehicle charging and offer customers alternatives as part of the evolving sharing economy.  Airbnb \cite{Airbnb} provides a peer-to-peer service as an alternative to traditional hotels, while Uber \cite{Uber} is a popular alternative to conventional taxi services.  The popularity of these services reveals a willingness of the public to use peer-to-peer alternatives, and this proposal will focus on the development of a distributed energy marketplace system.        

\begin{figure}[H]
    \centering
    \includegraphics[width=.9\linewidth, keepaspectratio]{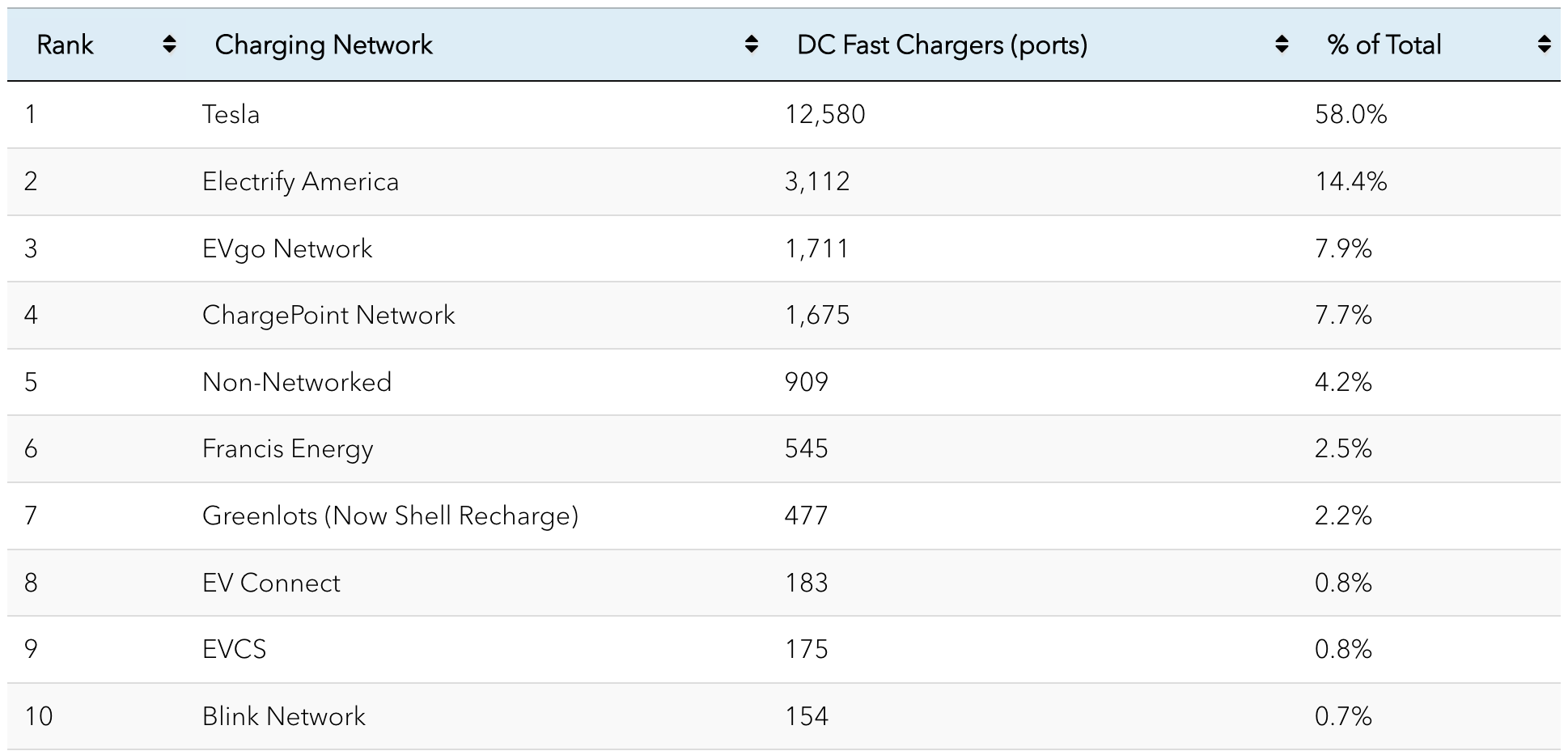}
    \captionsetup{width=.8\linewidth}
    % \decoRule
    \caption{Top 10 fast charging companies sorted by number of locations \cite{Evadaoption2022}.  The data was sourced from the US Department of Energy as of Dec 31, 2021 \cite{Doechargers2022}.}
    \label{fig:evadoption}    
\end{figure}

Although most of the charging can be done at home to facilitate local commutes, a publicly available network of chargers is still necessary for intercity commutes and to accommodate those who do not have charging facilities at home.  As a result of this demand, electric vehicle charging companies have rapidly formed and deployed charging networks in the United States.  Figure \ref{fig:evadoption} shows the top companies at the end of 2021 ranked by the total number of DC fast chargers.

\begin{figure}[H]
    \centering
    \includegraphics[width=.9\linewidth, keepaspectratio]{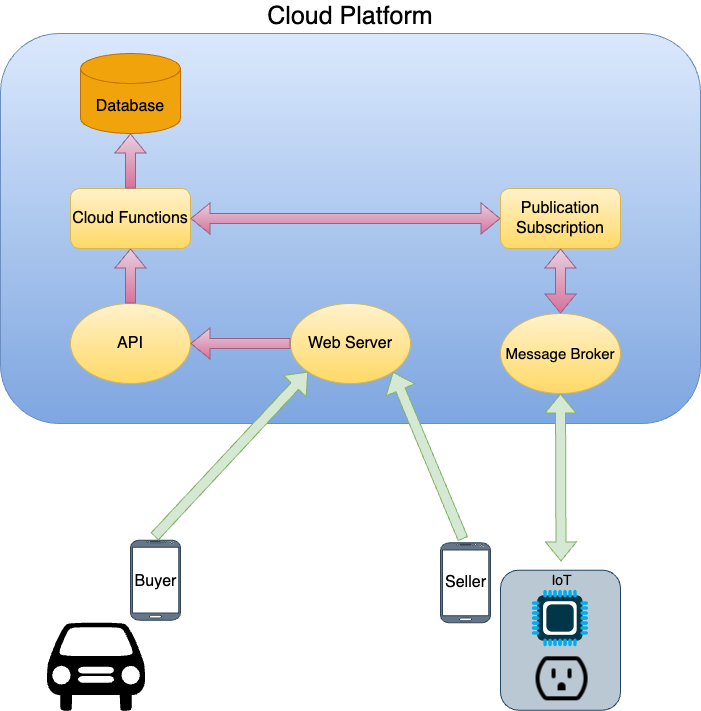}
    \captionsetup{width=.8\linewidth}
    % \decoRule
    \caption{A Cloud framework to implement the energy marketplace.  A seller registers their IoT device allowing buyers to search and reserve.}
    \label{fig:energy-marketplace}    
\end{figure}

Since the charging networks shown in Figure \ref{fig:evadoption} are still far from as ubiquitous as gas stations, we propose supplementing with private property owners who wish to rent their charging facilities through a peer-to-peer energy marketplace.  Figure \ref{fig:energy-marketplace} shows the architecture to implement the marketplace.  A seller will register their 240 volt level-2 charging outlet via an Internet of Things (IoT) microcontroller.  The IoT device communicates with a Message Broker within the Cloud framework, which is registered with a dedicated Publication-Subscription (PubSub) service.  PubSub provides a serverless synchronization mechanism between a message broker communicating with the IoT devices and the business logic running through Cloud Functions.  A Cloud Function may subscribe to notifications provided by the message broker representing a set of devices and also publish a new message to one or more devices. 

Both buyers and sellers interact with the marketplace through a web client.  Once a seller registers an IoT device, the Web API creates a corresponding database object.  All requests are processed through Cloud Functions.  The buyer will search for all chargers within a given latitude-longitude radius.  Within the search results, the buyer chooses the desired device, reserves it, and navigates to its location to start the charging session.  Once the session is completed, the device sends a detailed log that is stored in the database.  Additionally, Cloud Functions trigger payment processing via an external payment gateway (e.g. Stripe \cite{Stripe2023}).

A significant question in this design is what persistent storage needs are for this marketplace.  Our focus in this paper is the data flow caused by the aforementioned workflow and performance of the database.  At first, we do not know all the data storage requirements.  For example, when this framework is initially rolled out, only charger details such as:

\begin{lstlisting}[basicstyle=\ttfamily]
    name
    address
    latitude
    longitude
    type
    charging history: date, duration
    energy consumed
\end{lstlisting}\label{lst:initial-data}

\noindent will be persistently stored.  As development progresses, we will potentially add:

\begin{lstlisting}[basicstyle=\ttfamily]
    charger reservations and availability
    billing information
    user ratings and comments
    amenities
    nearby tourism information
    user profiles
\end{lstlisting}\label{lst:additional-data}

\noindent Given that the data fields and shape will evolve, a strict schematized database may cause additional administrator complexity, since existing data tables require a migration on a schema change.

Another concern is scalability.  If the marketplace has 10 sellers and 10 buyers, scaling is not an issue.  If it grows to 1,000,000 sellers, the database must be able to horizontally scale across multiple nodes with physical locations across multiple zones and regions.  A NoSQL database with a document model allows for this scaling with little administration.

Query latency and performance are also a concern.  Create, update, and delete operations may tolerate eventual consistency.  However, a read operation needs to query multiple fields, some within a range.  If a buyer searches for the level-2 charger type along with a latitude, longitude, and user rating range; a SQL query engine can easily handle this and return only matching records.  However, a NoSQL database such as Firestore will only be able to query a single field range at a time, requiring additional data reduction at the mobile application client.  

\subsection{Related Work}\label{sub:related-work}
The related work surveyed in this section provides an architecture and experimental evaluation background for the research topic of this paper.  We start with research on evaluating persistent storage for IoT device management.  The experimental setup, evaluation, and technologies used were valuable to guide our experiments and evaluation.  Our topic and our experiments use the surveyed research as a starting point without duplicating it.  Next, other research papers were surveyed that evaluated NoSQL databases against each other or against SQL.  These provide a methodology and benchmarking toolkit that is considered in the design of our experiments.

The next group of papers focuses on the specific technology used in our experiments.  Firestore, MySQL, and Spanner papers were surveyed and establish how performance is evaluated and the ability of each to scale.  Spanner provides a physical storage layer for the Firestore service and is thus included.  Finally, papers discussing approaches to augmenting MySQL so that it can scale were included.  These provide background to the challenges faced with horizontal scaling of an SQL database. 

\subsubsection{Performance Evaluation of IoT Data Management}\label{sub:rel-perf-iot}
Eyada et al. \cite{Eyada2020} is a paper focused on evaluating database latency and size as they are applied to large volumes of IoT sensor data.  Both the MySQL and MongoDB databases were evaluated while hosted in AWS EC2 instances. The experiments changed the workload, the compute resources, and the number of sensors.  The resulting latency values were fed into a predictive model equation developed with linear and nonlinear regression methods.  The results favor MongoDB for latency, while MySQL did not increase latency dramatically with increased sensor data.

\subsubsection{Benchmarking with YCSB}\label{sub-rel-ycsb}
Pandey \cite{Pandey2020} provides a comparison of relational and non-relational database platforms by comparing MySQL and MongoDB.  The benchmark suite Yahoo! Cloud Serving Benchmark (YCSB) was used to vary the workloads in each database and capture latency and throughput measurements.  All measurements favored the MongoDB solution.  However, the authors acknowledge and discuss the lack of strict atomicity, consistency, isolation, and durability (ACID) properties with non-relational, as well as the missing join operations for queries. The authors further conclude that MongoDB outperforms MySQL in sharding, security, performance, and availability.

Another paper discussing NoSQL platforms is by Khazaei et al. \cite{Khazaei2016}.  The authors explore some popular NoSQL databases and describe the characteristics of this solution.  The discussion includes the loosening of consistency, availability, partition tolerance (CAP) theorem, and the resulting basically available, soft-state, eventually consistent (BASE) systems.  The authors further compare multiple benchmarking suites including YCSB, PixMix, GRIDMix, and CALDA and conclude by choosing YCSB for its flexibility.   

\subsubsection{Google Firestore}\label{sub:rel-firestore}
The first paper written on the Firestore NoSQL database is by Kesavan et al. \cite{Kesavan2023} and is co-authored by a group of researchers within Google.  This paper digs into the architecture of Google's Firestore service, specifically addressing how it scales across many nodes, provides real-time notification capability, is easy to use, and provides a pay-as-you-go billing model.  The authors present benchmark data showing little increase in query latency given a large number of stored documents and increased document size and data shape.  They concluded that Firestore provides a convenient ecosystem with low barrier of entry for developers to rapidly prototype, deploy, iterate, and maintain applications.

Brito et al. \cite{Brito2019} discuss the key benefits and characteristics of GraphQL, a front-end programming interface and query language that wraps various data sources.  They then evaluated the benefits in practice by migrating seven systems.  The migration of the client APIs is discussed, and their resulting implementations reduced the returned data set by 94\% in terms of number of fields.

\subsubsection{Oracle MySQL}\label{sub:rel-mysql}
MySQL, a relational database, is widely used by most small and medium-sized applications \cite{Dong2015}.  It was initially released in 1995 and was developed by Oracle Corporation.  In Bannon et al. \cite{Bannon2002}, the authors dig into the MySQL server architecture and discuss the relational database management system (RDBMS) in generic terms that can be applied to other SQL systems.  The RDMBS is subdivided into application, logical, and physical layer software that runs directly on the node containing physical storage.  SQL requests are processed locally through a pre-compiler to extract the SQL statement, a query parser to convert to a parse tree structure, and an optimizer utilizing stored indices where possible.

\subsubsection{Sharding a Relational Database}\label{sub:rel-sharding}
Create, read, update, and delete (CRUD) operations are all supported through a Structured Query Language (SQL) interface.  These operations must adhere to a predefined schema that defines each column of the target table.  Schematization provides a sanity check to ensure that incoming row insertions are conforming to the desired data shape and column types.  Note that, under certain circumstances, it may be considered a restriction.  Any changes to a table's schema require the entire table to be migrated to the new shape.  Such a restriction may provide development hurdles as an application is rolled out iteratively and persistent data requirements evolve.

The support for SQL query processing is a benefit.  The language has been in existence since the early 1970s \cite{Chamberlin2012}, leading to a large user support group, extensive documentation, and teaching materials \cite{Taipalus2020}.  A huge advantage of the language is the ability to combine multiple inequality predicates within a single query.  For example, a \emph{ select} statement may contain multiple columns with $\ge$ conditions.  The statement is executed within a single operation and allows a response set to be minimized to only the desired matching records.  We will see later how this is significant to the energy marketplace.  SQL also supports \emph{join}.  Queries to different tables are combined within a single operation.  The \emph{join} allows data referenced through a relation to be accessed from their respective tables.

Although the external client may contain application and interface code, the remaining server architecture is largely monolithic and designed to run alongside physical storage within a single node.  As table sizes and concurrent requests grow, the owner organization will typically upgrade hosting hardware, i.e. faster processors, more memory, and additional disk storage, as described by Dong and Li \cite{Dong2015}.  The authors start their research by identifying the issue that many organizations have built persistent storage around a relational database management system (RDBMS) such as MySQL.  As the volume of information has seen an explosive increase, these systems are difficult to expand.  To address this, most organizations upgrade the database server hardware, migrate to a NoSQL or perform sharding such that the smaller partitions may be distributed.

The authors implemented a novel middleware application that allows multiple MySQL databases to interact as a single distributed entity.  A MySQL interface was exported to external clients.  Incoming queries were parsed and executed locally and then sent to the appropriate node in the distributed cluster of MySQL servers.  The middleware ran a sharding algorithm that allowed data slices to be sent to the appropriate data nodes.  A set of rules within the middleware configuration files defines the algorithm.  The master node implemented direct and semi-direct table join strategies to support queries across the distributed nodes.  After performing some functional tests, the authors concluded that their distributed database had a slower response time than a single database queried for the same data.  However, the advantage of their system is evident with massive datasets that cannot fit within a single MySQL server.

Yadav and Rahut \cite{Yadav2023} authored a paper that describes how Meta redesigned their MySQL datastore replication protocol to use a modified version of Raft \cite{Raft} instead of traditional semi-synchronous replication.  Raft is a consensus algorithm that uses an elected leader.  It imposes restrictions that only servers with the most up-to-date data can become leaders.  Data is sharded into many MySQL databases, utilizing a primary and many replicas.  Raft enables the control plane and data plane operations to be part of the same replicated log.  Membership and leadership are moved inside the MySQL server, resulting in provable correctness during promotions and membership changes.  This paradigm does not make sharding any easier, but rather optimizes the replication across multiple existing MySQL replicated databases.

\subsubsection{Google Spanner}\label{sub:rel-spanner}
Although Spanner is not used directly for the experiments in this paper, it is relevant since it provides the storage engine to Firestore.  Corbett et al. \cite{Corbett2013} is a collaboration of 23 authors within Google to discuss the mechanism and benefits of Spanner's distributed database service.  The authors continue to describe the architecture with a focus on distributed storage and timestamp management.  They claim that the database is ``semi-relational'' with schematized tables, yet it still supports an SQL query interface.  Below this layer, data is physically stored as key-value associated pairs across a large distributed network defined by a universe master, multiple zone masters and span servers.  The paper presents benchmark results for commit time of 1 to 200 participants (ranging from 14.6 to 122.5 milliseconds).  They conclude with a plug for the TrueTime timestamp API that is the ``linchpin'' of Spanner's advanced feature set.

%% file: Chapters/Chapter3.tex
\section{System Architecture}\label{sec:architecture}

\subsection{Firestore}\label{sub:firestore}
The first set of experiments is centered on a non-relational NoSQL database.  Firestore was chosen for its flexible billing structure and serverless architecture.  Firestore uses a document model \cite{Jatana2012} within its DBMS.  The first benefit of the document model is being schema-free.  Documents may be inserted with unstructured data, allowing changes in data shape (i.e. different fields and types) without requiring a migration of the whole database.  The unstructured format is especially helpful when an application such as the energy marketplace is first ramping up and the persistent data requirements may not be fully understood initially.  Thus, a schema-free paradigm allows incremental changes to what persisted as the application matures.  The unstructured format contributes to backward compatibility and database performance, since it does not need to be taken offline to perform migrations.

Another benefit of the document model is horizontal scaling.  Since the documents exist independently of each other, it is possible to distribute a collection across multiple physical compute nodes.  As document number and access traffic increases, this architecture allows for easy scaling with additional nodes.  Scaling is nearly linear and utilizes distributed hash tables (DHT) that organize pairs of [key, value] into storage buckets \cite{Pokorny2011}.  These buckets may span unlimited server nodes.

\begin{figure}[ht]
    \centering
    \includegraphics[width=1.0\linewidth, keepaspectratio]{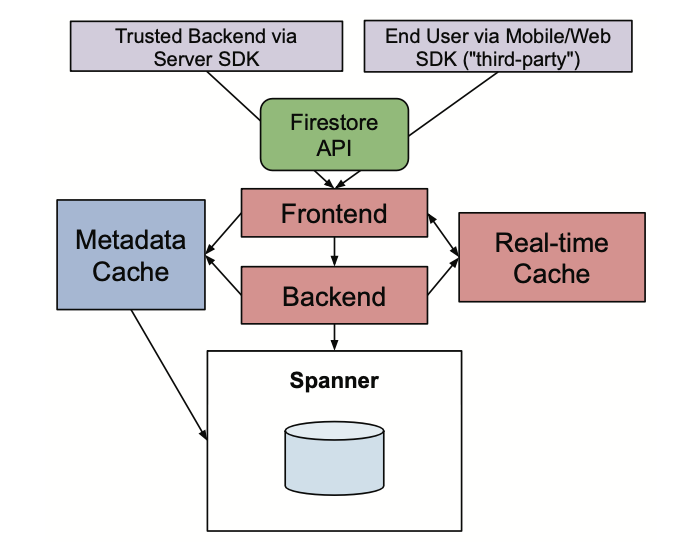}
    \captionsetup{width=.8\linewidth}
    % \decoRule
    \caption{Simplified diagram of the Firestore stack \cite{Kesavan2023}.}
    \label{fig:firestore-arch-stack}   
\end{figure}

Figure \ref{fig:firestore-arch-stack} shows a high-level overview of the Firestore stack.  Client access is provided through software development kit (SDK) libraries, which come in two flavors.  The ``Server'' SDK is trusted and used in applications running within a Google Cloud service such as a Compute VM or a Cloud Function.  It requires a privileged environment, foregoes authentication, and provides automatic retries with backoff.

The second SDK is ``Mobile and Web'', which is intended for untrusted third-party devices.  Both flavors of the SDK abstract the connection to the Firestore API.  They support blind writes and transactional writes based on optimistic concurrency control while connected.  Queries multiplex over the same long-lived connection to the Frontend task.  The Frontend tasks live in the same region as the database.  The co-existence means that client SDK requests arrive at the closest-to-the-user Google point of presence, then get routed to the Frontend task after the database location is fetched from the Firestore metadata.

\begin{figure}[H]
    \centering
    \includegraphics[width=1.0\linewidth, keepaspectratio]{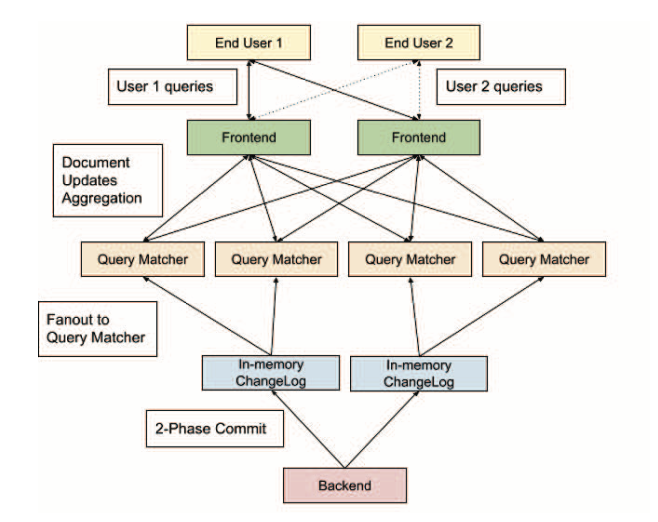}
    \captionsetup{width=.8\linewidth}
    % \decoRule
    \caption{Query processing through the Real-time Cache \cite{Kesavan2023}.}
    \label{fig:firestore-rt-cache}   
\end{figure}

After a Frontend task receives the SDK request, it sends a remote procedure call (RPC) to a Query Matcher.  Figure \ref{fig:firestore-rt-cache} expands on the Real-time Cache block from Figure \ref{fig:firestore-arch-stack} and shows the flow from multiple client users.  The In-memory ChangeLog ensures consistency through the check of timestamps.  Approved commits are forwarded to the Backend task.

Firestore executes all queries using secondary indices.  Each field in a document automatically generates an ascending and descending ordered index, all on a per-collection basis.  Automatic indexing may be excluded for a specific field if so configured.  Write operations are slightly more expensive latency wise, due to the need of updating multiple indices.  

Google Cloud Spanner \cite{Corbett2013} is used for the database storage engine.  Each Firestore database maps to a directory within some number of pre-initialized Spanner databases in the corresponding region.  Each directory has two tables, \emph{Entities} and \emph{IndexEntries} containing the Firestore database data.  Firestore documents are stored in the \emph{Entities} table, one document per row, with the contents (up to 1 megabyte) in a single column and the document key as the primary ID for the row.  Each index is stored within a single row of \emph{IndexEntries}.  To handle distribution, load balancing and scaling, Spanner performs automatic splitting and merging of rows into \emph{tablets}.  This process, similar to the sharding \cite{Cattell2011} performed by other relational database management systems (RDBMS), allows flexible horizontal scaling in an unlimited cluster of nodes.   

\subsection{MySQL}\label{sub:mysql}
As a relational database, data is stored in table objects that support relations between them, i.e. a data column in one table may point to a record in another table.  These relations support the normalization of data: A data item exists only in one table, yet may be referenced repeatedly from other tables in a one-to-many relationship.  Data normalization provides two main benefits. The first is that the database may take less physical storage because data duplication is not required. Second, write operations may require less latency since new data is only written to a single table and then referenced elsewhere if needed.

\begin{figure}[ht]
    \centering
    \includegraphics[width=1.0\linewidth, keepaspectratio]{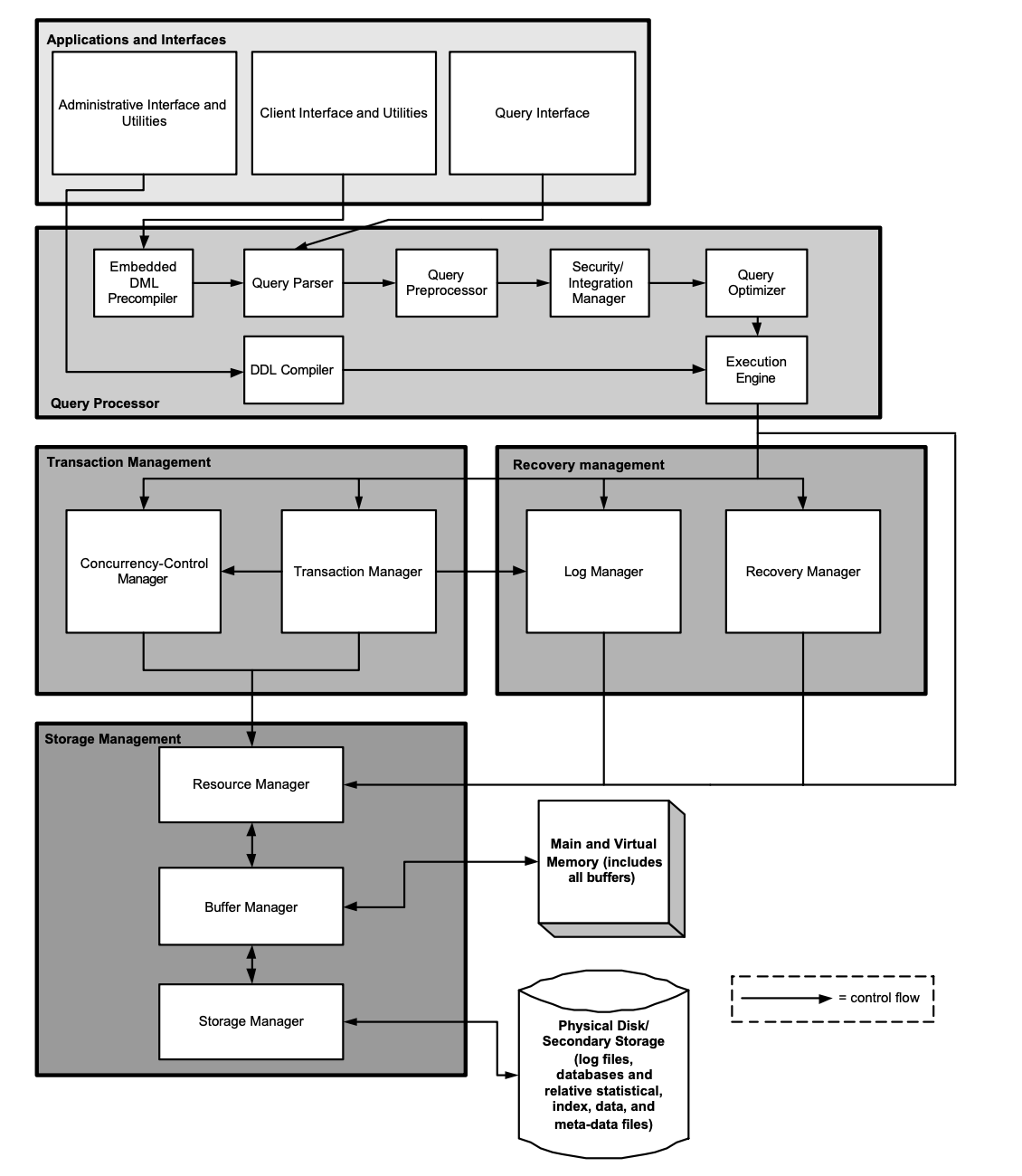}
    \captionsetup{width=.8\linewidth}
    % \decoRule
    \caption{Detailed heterogeneous conceptual MySQL architecture \cite{Bannon2002}.}
    \label{fig:mysql-arch}   
\end{figure}

Another benefit of the RDBMS model is the strict adherence to the ACID properties of a transaction.  These are atomicity, consistency, isolation, and durability.  The MySQL Transaction Management layer allows data manipulation operations that ensure that the database does not have the results of a partial operation \cite{Pandey2020}.  Figure \ref{fig:mysql-arch} graphically illustrates the Transaction Management sitting between the Query Processor and Storage Management.

\subsection{GraphQL}\label{sub:graphql}
GraphQL is not a product on its own, but rather a query-language specification.  The implementation used in this study is Apollo Server \cite{ApolloServer}, an open-source implementation compatible with any GraphQL client.  The server provides a middleware between the client and the database API.  It performs the following tasks on each query received:
\begin{enumerate}
    \item Parse: converts the query to an abstract syntax tree.
    \item Validate: checks that the query matches the schema.
    \item Execute: begins at the root of the tree and traverses, each field being a node in the tree.  Nodes at the same level execute their resolvers concurrently.
\end{enumerate}

All fields to query, variables passed in as query arguments, and response results are strongly typed.  GraphQL configuration involves defining an API: including defining types for each field (including condition arguments and return results) and the resolver functions that are correspondingly called when the field is included in the client query.

The GraphQL query structure is designed to resemble the expected results structure.  All desired values are returned with a single query of the GraphQL API.  The GraphQL server middleware then takes care of fetching the query fields from potentially multiple sources and endpoints.  Thus, the API acts as an aggregator that can greatly simplify the client complexity and provide a single endpoint enforced with strong typing for the client to send the query.   

\begin{figure}[ht]
    \centering
    \includegraphics[width=1.0\linewidth, keepaspectratio]{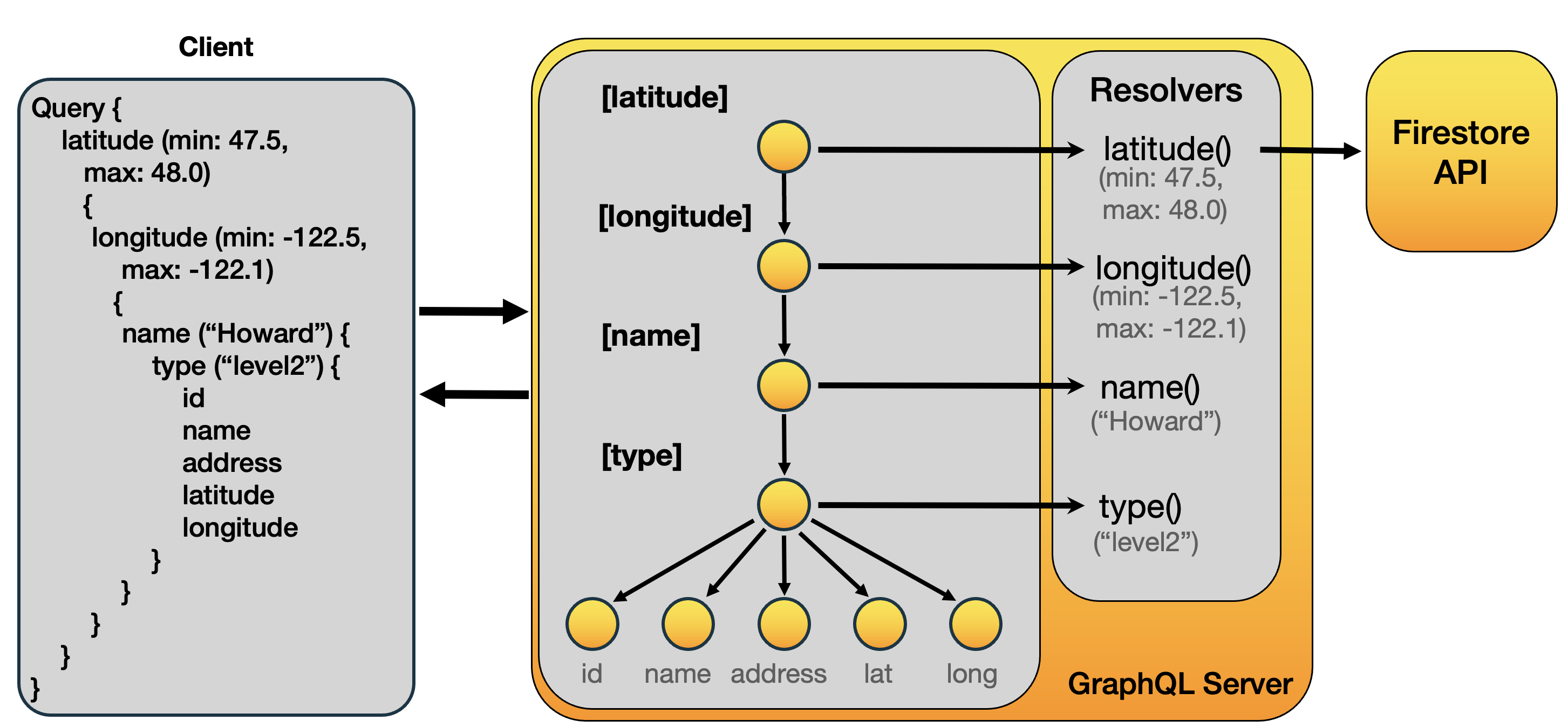}
    \captionsetup{width=.8\linewidth}
    % \decoRule
    \caption{Data tree and resolver sequence for a typical query as used in Section \ref{sub:exp-gql}.}
    \label{fig:gql-tree-resolver}   
\end{figure}

Figure \ref{fig:gql-tree-resolver} shows a query that is used in the experiments in Section \ref{sub:exp-gql}.  Rather than structuring the fields at the same level in the tree (which would cause the corresponding resolver functions to be called simultaneously), they are nested.  Nesting allows the resolver of each child field to apply to only the data set returned from its parent.  Since Firestore cannot support multiple field range queries, Figure \ref{fig:gql-tree-resolver} demonstrates how the top field in the GraphQL query calls the Firestore API, but subsequent child field resolvers only filter the parent field's data.  Since GraphQL abstracts the back-end database, there exists flexibility to switch from NoSQL to SQL (or any other persistent storage solution) in the future. 

%% file: Chapters/Chapter4.tex
\section{Experiments}\label{sec:experiments}
The experiments provide a comparative benchmark of read, update, and create performance between relational and non-relational database engines.  In all cases, $n$ represents the number of documents in a non-relational collection or the records in a relational table.  The set of documents or records (chargers) in the response is represented by $R$ and $r$ is the size (number of chargers) of $R$.  Instead of using a third-party benchmarking tool, these experiments use a custom client that performs read, update, and create operations that mimic workflows in the energy marketplace.  The registration of a new IoT device, reading its persistent data, and searching for specific combinations of fields are shown in Figure \ref{fig:energy-marketplace-workflow} and are covered in these experiments.  In addition to performing operations to mimic the marketplace workflow, we also hit the databases with high-volume operations to determine the limits of read-data rate and read-create operation rate.

The client, as described in Figure \ref{fig:energy-marketplace}, is a mobile Edge device that sends request messages to the Cloud framework.  However, for the purposes of the experiments described in this section, we used a workstation-based custom client.  This client emulates a user sending requests through their mobile device and encapsulates all the database library code for both Firestore and MySQL, as well as the workflow scenarios.  All of the following experiments that perform iterative operations do so sequentially.  Both the iterative read and iterative create experiments wait for each operation to complete prior to executing the next.  The sequential execution allows the characterization of the read and create operation rate, but does not take into account contention from multiple clients.  In a production energy marketplace scenario, read and update requests may be coming in from many clients simultaneously, which we do not characterize in these experiments. 

\subsection{Non-Relational, NoSQL Database}\label{sub:exp-non-rel}
The test database is divided into six collections ranging from $10^1$ to $10^6$ documents.  These documents represent mock data for the energy marketplace.  Each is a sample registration of a charger device with \emph{name}, \emph{id} number, \emph{latitude}, \emph{longitude}, \emph{address} and charger \emph{type}.  All have a unique key string in addition to the listed fields.  Each mock data set is generated with a random generator algorithm that cycles each field between a range of accepted values.  The fields \emph{latitude}, \emph{longitude}, and \emph{name} cycle between five values, the \emph{address} cycles through four street names with random house numbers, while the \emph{type} cycles between two levels. The \emph{latitude} values range from $43.1$ to $47.9$ while the \emph{longitude} ranges from $-124.9$ to $-120.1$.  Table \ref{tab:mock-data} shows the range of all fields, including cardinality.  The smaller the cardinality number, the less unique the values in its distribution.  Thus, the \emph{type} field with a cardinality of one has a high level of repetition.  The client tests update performance through a single update to each collection (i.e. reserving a new charger), followed by rapid sequential creates of all documents.  These experiments measure both the time it takes to execute a single update and how many create operations per second are supported by the database.  Read performance is measured through a single document read of each collection followed by a rapid sequential single read of all documents.  The time to execute a single read operation is measured followed by the attainable read operations per second.  Next, the client performs a bulk read of all documents in each collection.  The time to execute provides the data rate that can be read.

\begin{table}[h]
    \captionsetup{width=.8\linewidth}
    \caption{The mock data distribution.  The data generator creates sets of mock data, one for each $n$.  The $id$ field sequentially increases for every mock charger.  The $name$ field has a set of five names which are randomly selected. 
    The $address$ field randomly selects a house number the given range and then randomly selects one of four street names.  The $latitude$ and $longitude$ randomly select between the range.  The $type$ randomly selects between two. }
    \label{tab:mock-data}
    \centering
    \begin{tabular}{p{0.2\linewidth} p{0.4\linewidth} p{0.2\linewidth}}
    \toprule
    \tabhead{Field} & \tabhead{Range} & \tabhead{Cardinality}\\
    \midrule
    $id$ & 1-$n$ & 996,402\\
    $name$ & Howard, Gomez, Singh, Shipman, Durnin & 4 \\
    $address$ & 10000-11900: Cedar Ct, 118th Ave, 119th Ave Maple Hill Ln & 7,341 \\
    $latitude$ & 43.19 - 47.9 & 4 \\
    $longitude$ & -124.9 - (-120.1) & 4 \\
    $type$ & level1 or level2 & 1\\
    \bottomrule\\
    \end{tabular}
\end{table}

Finally, pattern querying measures the combinations of reading.  We define a specific query pattern with equality conditions for the \emph{name} and \emph{type} fields and inequality ranges for both \emph{latitude} and \emph{longitude}.  Two queries execute against each collection:
\begin{itemize}
    \item Read all documents and search for matches.
    \item Read documents matching a single field range query (\emph{latitude}) and search the subset for remaining matches.
\end{itemize}

\subsubsection{Reads}\label{sub:nonrel-reads}
\begin{figure}[H]
    \centering
    \includegraphics[width=1.0\linewidth, keepaspectratio]{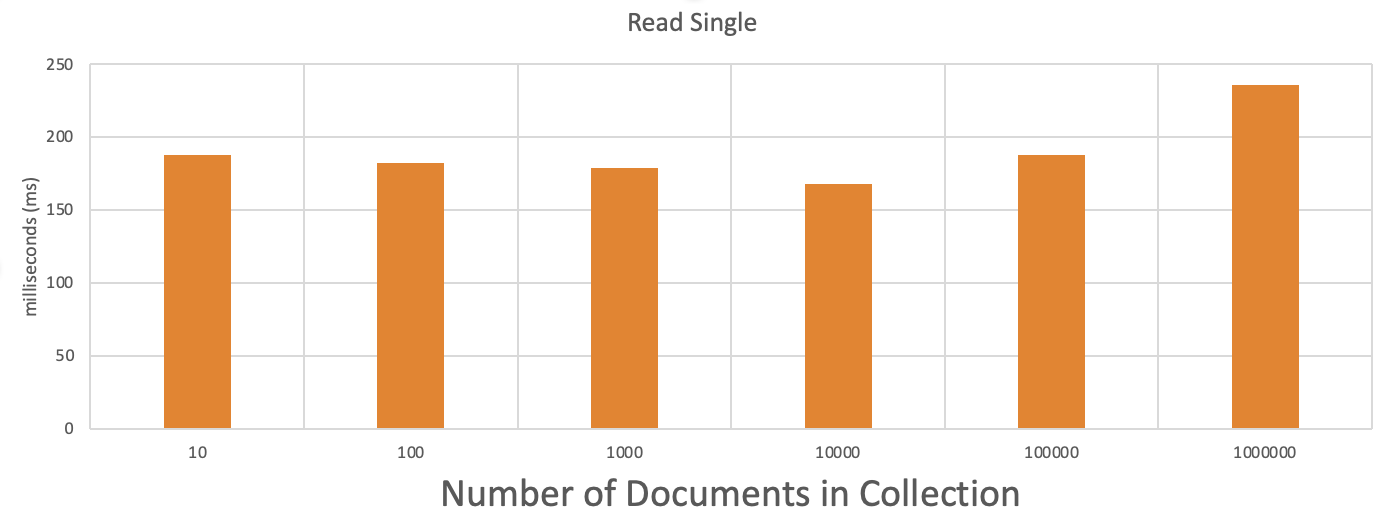}
    \captionsetup{width=.8\linewidth}
    % \decoRule
    \caption{A single read operation performed against each of the 6 collections.  The time to complete the operation is shown in milliseconds (ms).}
    \label{fig:firestore-read-single}   
\end{figure}

For the first experiment, a single document was read from each collection using the document key (primary index).  The 6 collection sizes ($n$) are:
\begin{enumerate}
    \item $n$ = 10 documents
    \item $n$ = 100 documents
    \item $n$ = 1000 documents
    \item $n$ = 10000 documents
    \item $n$ = 100000 documents
    \item $n$ = 1000000 documents
\end{enumerate}
For each, the document key \emph{1\_11899NW118thAve} was used to perform a single read operation.  Figure \ref{fig:firestore-read-single} shows that the time to complete each read operation is reasonably consistent across the collection size range.  The distributed hash table (DHT) lookup of the primary key explains this consistency.

The next experiment involves iteratively reading every document in each collection.  The \emph{id} field, a secondary index, is used to perform a simple read operation.  The \emph{id} field is a sequential integer assigned to each document.  It has no direct meaning for the charger represented in the document, but allows for an easy sequential query of all documents in the collection.  Note that this field is automatically indexed by the Firestore service to increase the performance of their query engine.

\begin{figure}[h]
    \centering
    \includegraphics[width=1.0\linewidth, keepaspectratio]{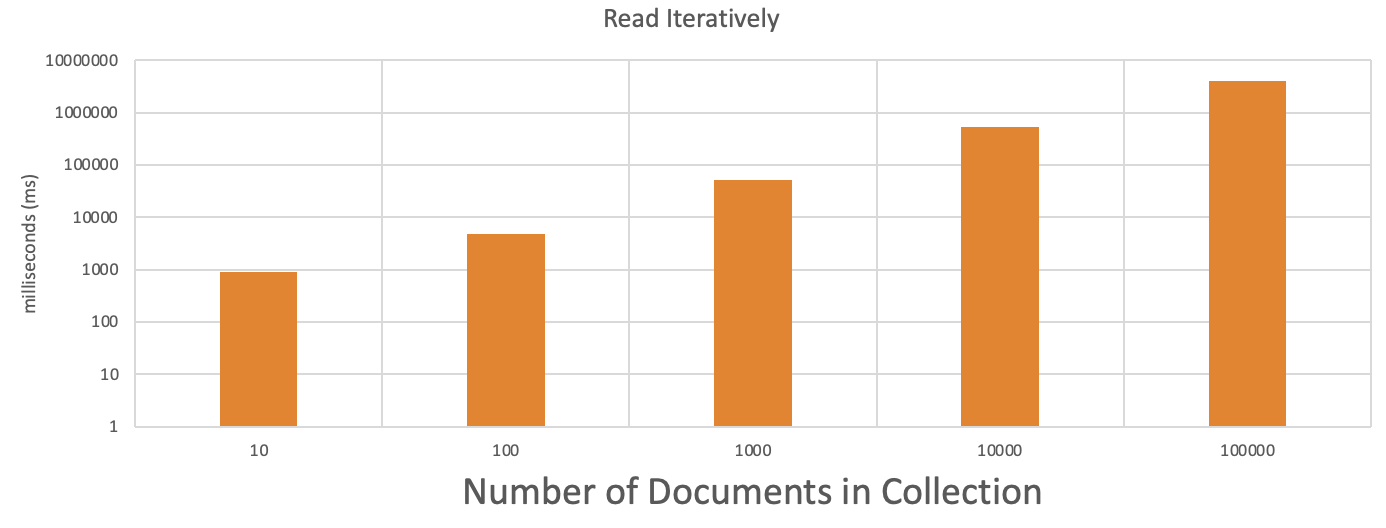}
    \captionsetup{width=.8\linewidth}
    % \decoRule
    \caption{Iterative read operations performed against each of the 6 collections using a secondary index.  Each unique \emph{id} field is successively queried. Time to complete all the operations is shown in milliseconds (ms) on a logarithmic scale.  There were repeated system timeouts while iteratively querying the $10^6$ collection, thus we only show five collections here.}
    \label{fig:firestore-read-iterative}   
\end{figure}

Figure \ref{fig:firestore-read-iterative} shows the time taken to iteratively read all the documents.  A sequential read like this does not directly represent a user workflow in the marketplace, but allows us to characterize the read operation rate of the database (shown in Table \ref{tab:firestore-read-metrics}).  Time complexity consists of the query time for each \emph{id} field across the $n$ documents in the collection.  Thus, the total time is
\begin{equation}\label{eq:iter-time1}
    T_{C}(n) = O(n + O_{D}(n))
\end{equation}
where \\
$D$: A single read operation via an \emph{id} query.\\
$C$: Iteration across the entire collection.\\

Since the query of each document utilizes a pre-generated secondary index, the query engine performs a binary search to quickly match the \emph{id} value in the index.  This binary search has the worst-case time complexity of $T(n) = O(\log{n})$ for the first lookup and sequential after that.  Substituting this into Equation \ref{eq:iter-time1}, we can simplify to:
\begin{equation}
     T(n) = O(n + \log{n}) = O(n)
\end{equation}
The time complexity is linear.  Note that $r = 1 \forall n$ since every iteration returns exactly one document.

\begin{figure}[h]
    \centering
    \includegraphics[width=1.0\linewidth, keepaspectratio]{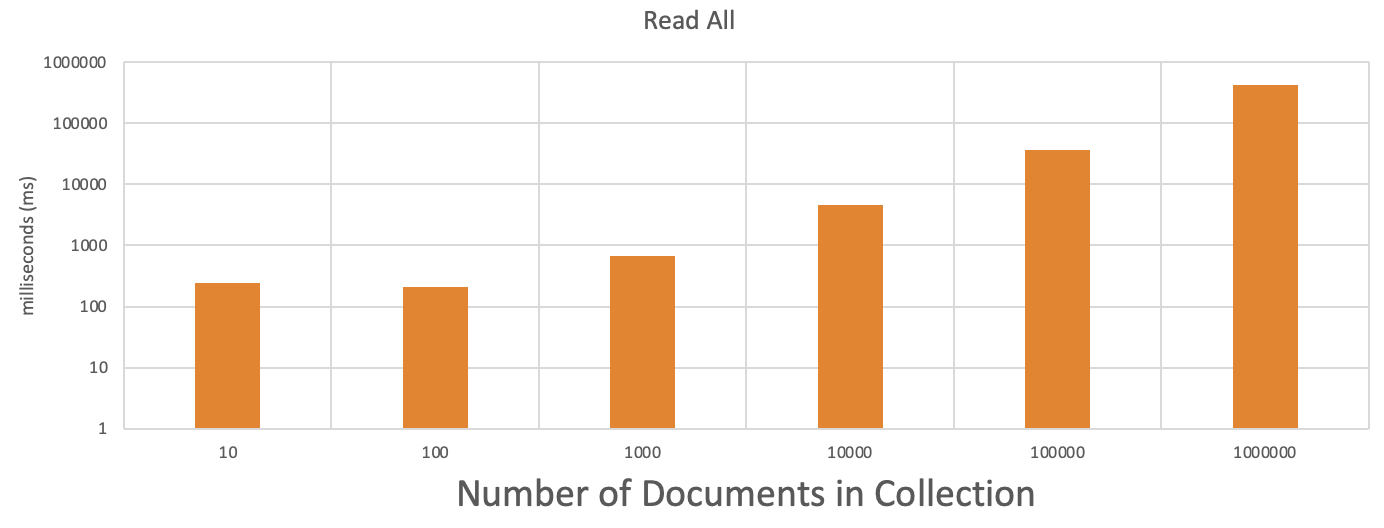}
    \captionsetup{width=.8\linewidth}
    % \decoRule
    \caption{A read all operation performed against each of the 6 collections.  All documents are read in a single operation. Time to complete all the operations is shown in milliseconds (ms) on a logarithmic scale.}
    \label{fig:firestore-read-all}   
\end{figure}

The final read experiment involves a read of all the documents in each collection.  The operation does not use a primary or secondary index.  Figure \ref{fig:firestore-read-all} shows the time taken to read all documents by collection.  In this scenario, $r = n \forall n$ since all documents in the collection are always returned.

The time to read all documents is directly proportional to the number of documents returned. A bulk query such as this is not in any user workflow, but allows characterization of the data rate as seen in Table \ref{tab:firestore-read-metrics}. The time complexity for this operation is linear and is represented by Equation \ref{eq:read-all-time}.
\begin{equation}\label{eq:read-all-time}
    T(n) = O(n)
\end{equation}

We finish the read analysis with the performance metrics shown in Table \ref{tab:firestore-read-metrics}.  The time in milliseconds to perform a single read operation with a known key (primary index) is taken directly from Figure \ref{fig:firestore-read-single}.  Operations per second are calculated by dividing the number of iterative operations in Figure \ref{fig:firestore-read-iterative} by the time to complete. The kilobytes per second data rate uses an average document size of 76 bytes, multiplies by the number of documents, and divides by the time taken to complete a read-all operation as shown in Figure \ref{fig:firestore-read-all}. 

\begin{table}[h]
    \captionsetup{width=.8\linewidth}
    \caption{A summary of all read metrics.  The first column represents the number of documents in the collection.  The second is the time taken for a single read operation with a known key.  The third is the operations performed per second while iteratively reading all documents via a secondary key.  The fourth column represents data rate during a read-all operation.}
    \label{tab:firestore-read-metrics}
    \centering
    \begin{tabular}{l r | l l l}
    \toprule
    \tabhead{} & \tabhead{n} & \tabhead{ms/oper} & \tabhead{ops/s} & \tabhead{kB/s} \\
    \midrule
    & 10 & 188 & 10.917 & 9.708 \\
    & 100 & 182 & 20.593 & 109.390 \\
    & 1,000 & 179 & 19.247 & 340.643 \\
    & 10,000 & 168 & 19.250 & 503.893 \\
    & 100,000 & 188 & 24.759 & 644.269 \\
    & 1,000,000 & 236 & & 554.681 \\
    \midrule
    \tabhead{Mean} & & 190.167 & 18.953 & 360.420 \\
    \bottomrule\\
    \end{tabular}
\end{table}

\subsubsection{Updates and Creates}\label{sub:nonrel-writes}
For the next set of experiments, we will focus on characterizing the update and create operations.  These largely mirror Section \ref{sub:nonrel-reads} and much of the background explanation applies.  The first experiment performs a single update operation on an existing document in each collection.  A known key, \emph{1\_11899NW118thAve}, is used in all cases.  Note that for all create operations, a previously generated and randomized set of mock data is used to represent each charger device.

\begin{figure}[h]
    \centering
    \includegraphics[width=1.0\linewidth, keepaspectratio]{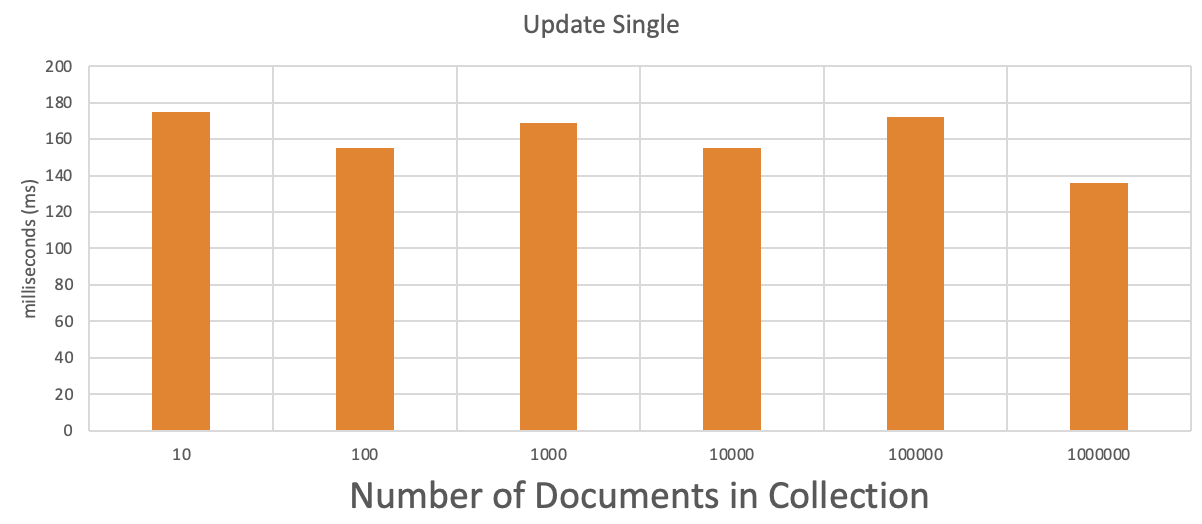}
    \captionsetup{width=.8\linewidth}
    % \decoRule
    \caption{A single update operation performed against each of the six collections.  Time to complete the operation is shown in milliseconds (ms).}
    \label{fig:firestore-update-single}
\end{figure}

\begin{figure}[h]
    \centering
    \includegraphics[width=1.0\linewidth, keepaspectratio]{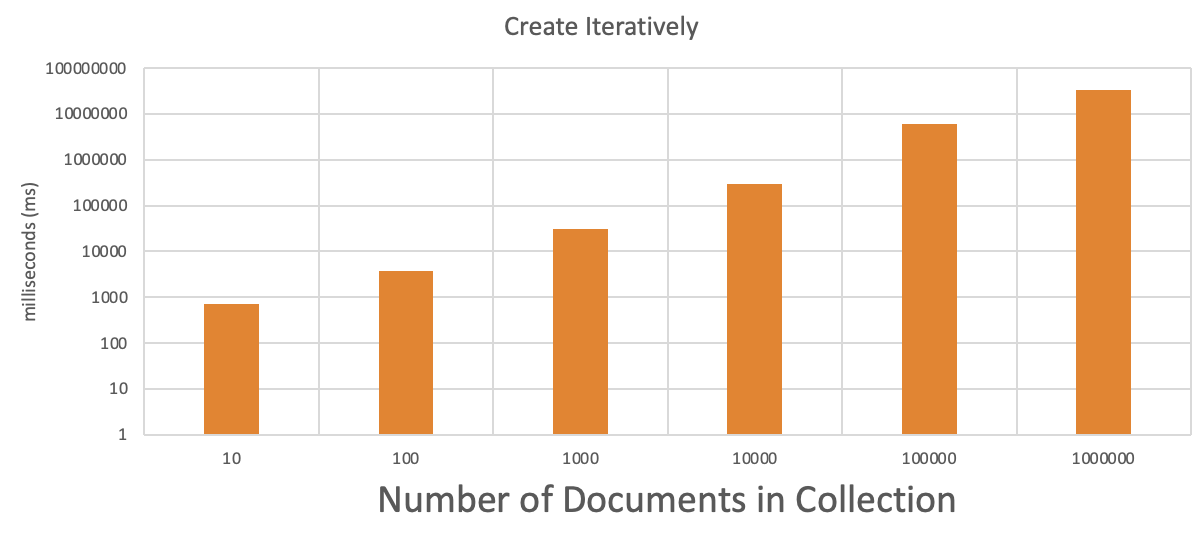}
    \captionsetup{width=.8\linewidth}
    % \decoRule
    \caption{An iterative create operation performed against each of the 6 collections.  All $n$ documents are created in sequence. Time to complete all the operations is shown in milliseconds (ms) on a logarithmic scale.}
    \label{fig:firestore-create-iterative}   
\end{figure}

Figure \ref{fig:firestore-update-single} shows the time in milliseconds for the operations to complete.  As with reads, a single update completes in consistent time regardless of the size of the collection.  Next, Figure \ref{fig:firestore-create-iterative} shows the iterative create of all documents in the collections.  As with reads, this scenario does not represent a user workflow, but allows us to characterize the create operation rate, as shown in Table \ref{tab:firestore-write-metrics}.  A high create operation rate is not critical for the energy marketplace, but is a useful metric when comparing NoSQL vs. SQL.  The time complexity for all $n$ operations to complete follows that of the read scenario where $T(n) = O(n + \log{n})$ or linear time.  Note that a create-all operation is not possible through this client interface and is not tested here.

\begin{table}[H]
    \captionsetup{width=.8\linewidth}
    \caption{A summary of the update and create metrics.  The first column represents the number of documents in the collection.  The second is the time taken for a single update operation with a known key.  The third is the operations performed per second while iteratively creating all documents.}
    \label{tab:firestore-write-metrics}
    \centering
    \begin{tabular}{l r | l l}
    \toprule
    \tabhead{} & \tabhead{n} & \tabhead{Update ms/oper} & \tabhead{Create ops/s} \\
    \midrule
    & 10 & 175 & 13.755 \\
    & 100 & 155 & 27.211 \\
    & 1,000 & 169 & 31.625 \\
    & 10,000 & 155 & 33.789 \\
    & 100,000 & 172 & 16.469 \\
    & 1,000,000 & 136 & 29.448 \\
    \midrule
    \tabhead{Mean} & & 160.333 & 25.383 \\
    \bottomrule\\
    \end{tabular}
\end{table}

Table \ref{tab:firestore-write-metrics} summarizes the milliseconds per update operation as shown in Figure \ref{fig:firestore-update-single} and the create operations per second that are possible as shown in Figure \ref{fig:firestore-create-iterative}.  Although the update and create results largely mirrored the read, we see a slight performance improvement. 

\subsection{Reads with Conditions}\label{sub:nonrel-queries}
The set of experiments that most mirrors real-world scenarios is the querying of complex field patterns.  Multiple range queries (inequality condition) of secondary indices are not allowed in Firestore. Therefore, to read the documents that match all the query conditions, some combination of querying the database and data reduction at the client must be performed.  For these experiments, the following query conditions are used:

\begin{lstlisting}[basicstyle=\ttfamily]
    name: Howard
    latitude: min 47.5 and max 48.0
    longitude: min -122.5 and max -122.1
    type: level2
\end{lstlisting}\label{lst:query-parms}

\noindent The first of the query experiments reads all documents without a query.  The client is an Edge device coded in JavaScript that utilizes the Firestore development kit.  Client-side searching is performed to narrow down the documents in each collection that match the given query parameters.  This is the most brute-force method and is shown here for comparison.  Since we always receive every document, $\forall n \; r = n$.  Also, the search performance on the client side will have some variability due to the processor capability.

\begin{figure}[H]
    \centering
    \includegraphics[width=1.0\linewidth, keepaspectratio]{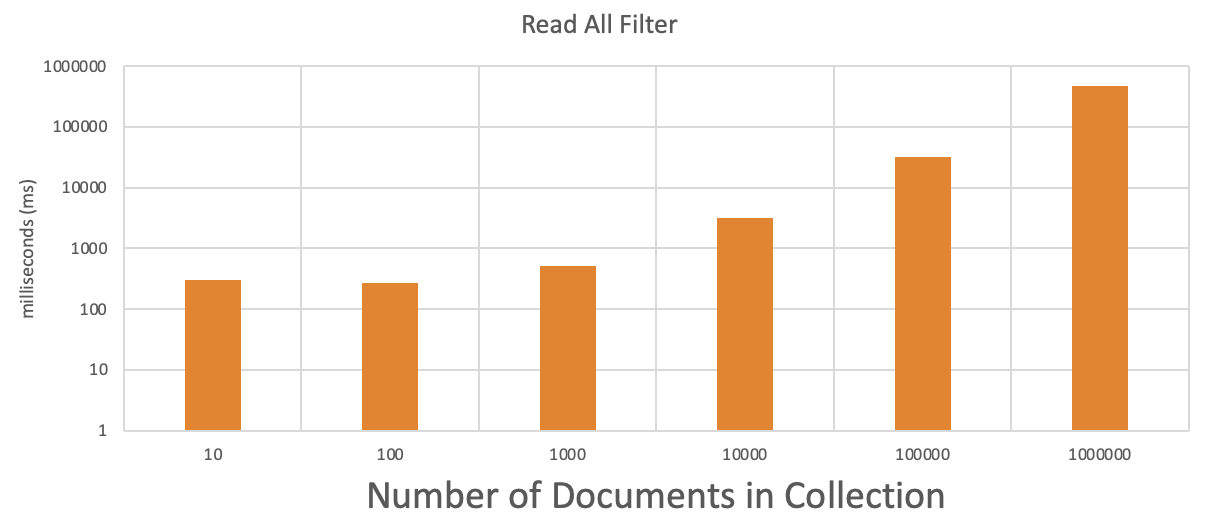}
    \captionsetup{width=.8\linewidth}
    % \decoRule
    \caption{A read all performed against each of the 6 collections.  All documents are read in a single operation, then reduced locally to those matching the query conditions. Time to complete all the operations is shown in milliseconds (ms) on a logarithmic scale.}
    \label{fig:firestore-read-all-filter}   
\end{figure}

Figure \ref{fig:firestore-read-all-filter} shows the time to read all documents and reduce matches very closely with the time to read all from Figure \ref{fig:firestore-read-all}.  The reduction time on the test machine (M1 processor) is negligible.  Thus, the response set size $r$ is the defining factor for time complexity, which is linear time or $T(n) = O(n)$.

\begin{figure}[H]
    \centering
    \includegraphics[width=1.0\linewidth, keepaspectratio]{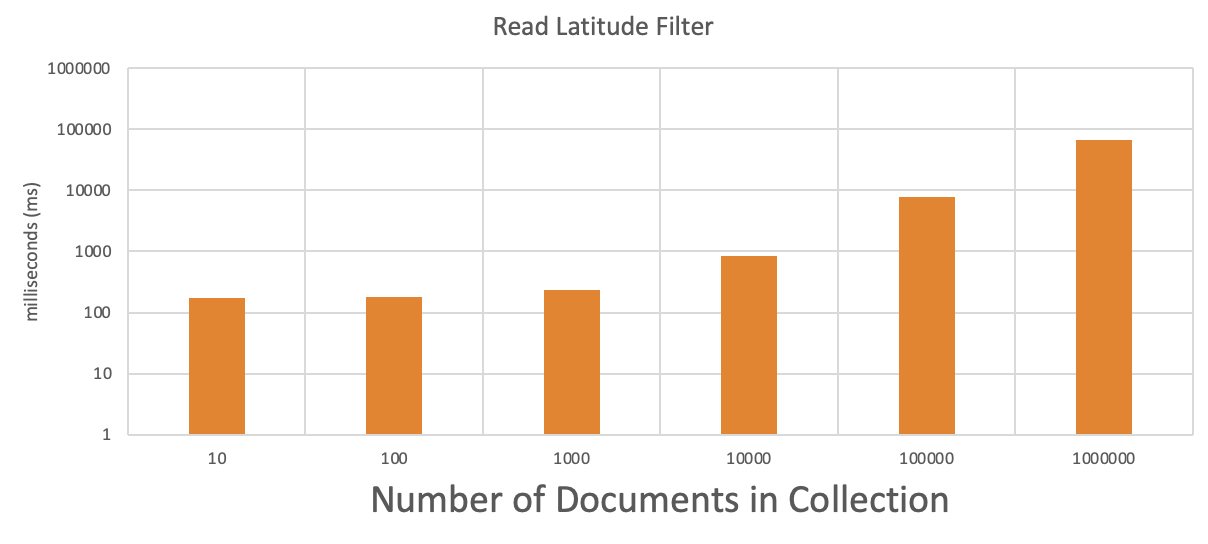}
    \captionsetup{width=.8\linewidth}
    % \decoRule
    \caption{Query the \emph{latitude} range in each of the 6 collections.  All \emph{latitude}-matching documents are read in a single operation, then reduced locally to those matching the non-\emph{latitude} query conditions. Time to complete all the operations is shown in milliseconds (ms) on a logarithmic scale.}
    \label{fig:firestore-read-lat}   
\end{figure}

The next experiment performs a conditional query using the \emph{latitude} field.  It defines a range between the minimum and maximum values. The Firestore query engine can only handle a single field range; thus we query only with \emph{latitude}.  Firestore performs a binary search on the pre-generated \emph{latitude} index and returns an inflated $r$ set of documents that must be searched locally on the client to match \emph{name}, \emph{type} and \emph{longitude}. Figure \ref{fig:firestore-read-lat} shows the time spent querying and subsequent data reduction.    

As expected, querying an index and reducing the smaller $r$ yields lower latency times than querying all documents to reduce.  Table \ref{tab:firestore-conditional-read-comparison} shows the reduction in latency achieved by the conditional read of \emph{latitude}, followed by the data reduction.

\begin{table}[h]
    \captionsetup{width=.8\linewidth}
    \caption{Summary of conditional read latency.  The first column is the collection queried.  The second column represents the latency required to read all documents with data reduction.  The third reads only the documents matching the \emph{latitude} condition with data reduction.  The fourth represents the latency difference between the two. }
    \label{tab:firestore-conditional-read-comparison}
    \centering
    \begin{tabular}{r | R{0.2\linewidth} R{0.2\linewidth} R{0.2\linewidth}}
    \toprule
    \tabhead{n} & \tabhead{Read All \newline Filter (ms)} & \tabhead{Read Latitude Filter (ms)} & \tabhead{Difference (ms)} \\
    \midrule
    10 & 300 & 171 & -129 \\
    100 & 273 & 182 & -91 \\
    1,000 & 508 & 234 & -274 \\
    10,000 & 3,213 & 855 & -2,358 \\
    100,000 & 31,425 & 7,718 & -23,707 \\
    1,000,000 & 459,787 & 66,763 & -403,024\\
    \bottomrule
    \end{tabular}
\end{table}

\subsection{Relational, SQL Database}\label{sub:exp-rel}
The next set of experiments focuses on the SQL database and how the performance differs from NoSQL.  The terminology we use to describe the experiments changes slightly from the NoSQL scenario in Section \ref{sub:exp-non-rel}.  ``Collections'' and ``documents'' are now replaced with ``tables'' and ``records'' (or ``rows'').  ``Fields'' are replaced with ``columns''.  The same mock data from our NoSQL experiments is inserted into 6 tables that span $10^1$ to $10^6$ records.  The largest difference in the experimental procedure is that queries always return exactly the matching results.  There is no need for the scenario described in Section \ref{sub:nonrel-queries} where a larger-than-needed result size $r$ is returned and further reduced client-side to the matching records.

\subsubsection{Reads, Updates and Creates}\label{sub:rel-reads-writes}
The single and sequential read, update, and create operations show similar results to Firestore in Section \ref{sub:exp-non-rel}.  We only summarize.  A summary of the read metrics, Table \ref{tab:mysql-read-metrics}, shows that milliseconds per operation and operations per second remain fairly constant with increasing $n$.  However, the data rate increases proportionally to increasing $n$ due to the almost constant read time. 

\begin{table}[H]
    \captionsetup{width=.8\linewidth}
    \caption{A summary of time per operation, operations per second and data rate in kilobytes for each of the 6 table sizes.  The ops/s values are derived from the iterative read experiments while kB/s comes from the read all. }
    \label{tab:mysql-read-metrics}
    \centering
    \begin{tabular}{l r | l l l}
    \toprule
    \tabhead{} & \tabhead{n} & \tabhead{ms/oper} & \tabhead{ops/s} & \tabhead{kB/s} \\
    \midrule
    & 10 & 84 & 38.911 & 26.477 \\
    & 100 & 82 & 54.975 & 274.118 \\
    & 1,000 & 105 & 55.630 & 1,879.032 \\
    & 10,000 & 86 & 63.180 & 12,944.444 \\
    & 100,000 & 90 & 62.519 & 48,240.166 \\
    & 1,000,000 & 106 & 48.751 & 624,664.879 \\
    \midrule
    \tabhead{Mean} & & 92.167 & 53.994 & 114,671.519 \\
    \bottomrule\\
    \end{tabular}
\end{table}

\begin{table}[H]
    \captionsetup{width=.8\linewidth}
    \caption{A summary of time per operation (update) and operations per second (create) for each of the 6 table sizes.}
    \label{tab:mysql-write-metrics}
    \centering
    \begin{tabular}{l r | l l}
    \toprule
    \tabhead{} & \tabhead{n} & \tabhead{Update ms/oper} & \tabhead{Create ops/s} \\
    \midrule
    & 10 & 101 & 26.385 \\
    & 100 & 80 & 52.329 \\
    & 1,000 & 88 & 49.317 \\
    & 10,000 & 95 & 53.332 \\
    & 100,000 & 96 & 53.975 \\
    & 1,000,000 & 91 & 38.955 \\
    \midrule
    \tabhead{Mean} & & 91.833 & 45.715 \\
    \bottomrule\\
    \end{tabular}
\end{table}

We complete the summary of the update and create metrics in Table \ref{tab:mysql-write-metrics}.  The average create operation rate was approximately 46 operations per second regardless of the size of the table $n$.  With both the update time per operation and the create operations per second approximately constant, we conclude that both updates and creates to a SQL, relational database may be performed in constant time or $T(n) = O(1)$.

\subsubsection{Reads with Conditions}\label{sub:rel-queries}
\begin{figure}[H]
    \centering
    \includegraphics[width=1.0\linewidth, keepaspectratio]{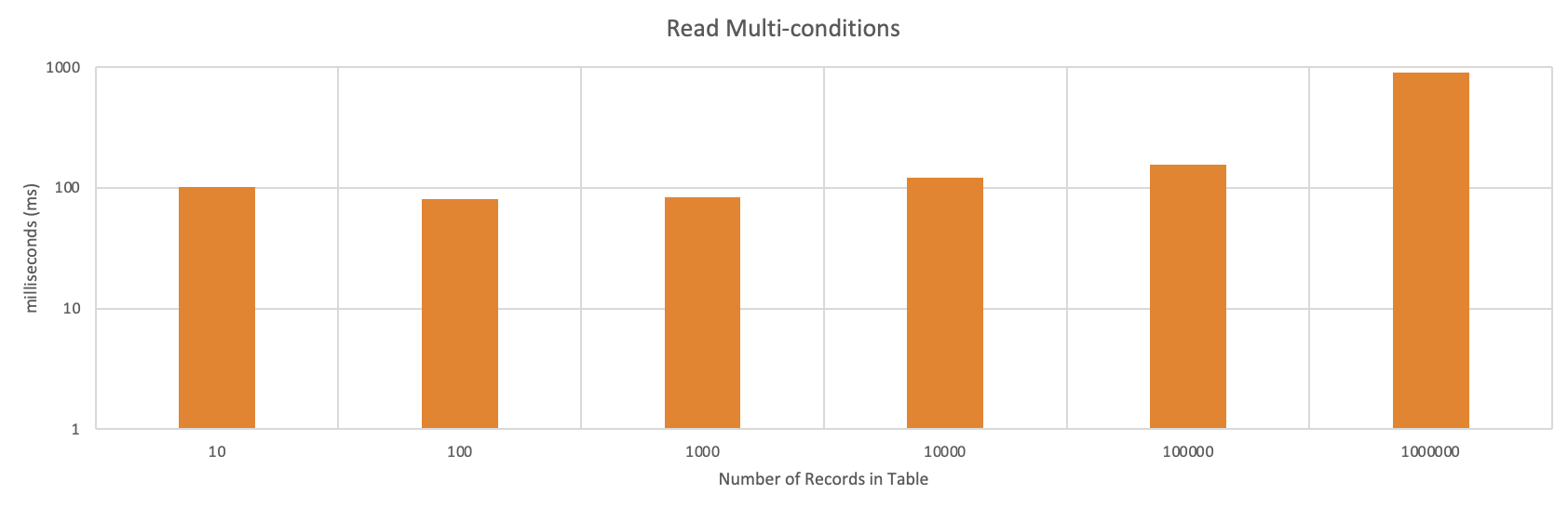}
    \captionsetup{width=.8\linewidth}
    % \decoRule
    \caption{Query with name=Howard, type=level2, 47.5 $\le$ latitude $\le$ 48.0, -122.5 $\le$ longitude $\le$ -122.1.  Time to complete all the operations shown in milliseconds (ms) across all 6 tables.}
    \label{fig:mysql-read-conditions}   
\end{figure}

The final set of benchmarks is based on the same conditions listed in Section \ref{sub:nonrel-queries}.  These results are the most significant, since they show the scenario in which the SQL database outperforms the NoSQL.  This experiment represents the marketplace scenario where a user queries a \emph{ latitude, longitude} range along with charger \emph{type} and seller \emph{name} as conditions. Figure \ref{fig:mysql-read-conditions} shows a very slight increase in completion time as the size of the table $n$ increases.  The increase in time can be attributed to the larger response size $r$.  Note that, unlike with the non-relational, multi-field range queries, only the matching records are returned.   As the response set grows larger (i.e. $n = 10^6$), the execution time increases slightly.  The increase is proportional to $r$, but is kept minimal since only records that match the query conditions are returned.

The default behavior of MySQL is to not generate indexes for each column in addition to the PRIMARY key.  The lack of auto-generated indexes was confirmed both on the Cloud SQL instance and on a local MySQL test server.  The \emph{SHOW INDEX} and \emph{EXPLAIN} SQL commands confirm that no indexes were either generated or used during the course of executing the queries in the experiments above.

\subsection{GraphQL with NoSQL}\label{sub:exp-gql}
\begin{figure}[H]
    \centering
    \includegraphics[width=1.0\linewidth, keepaspectratio]{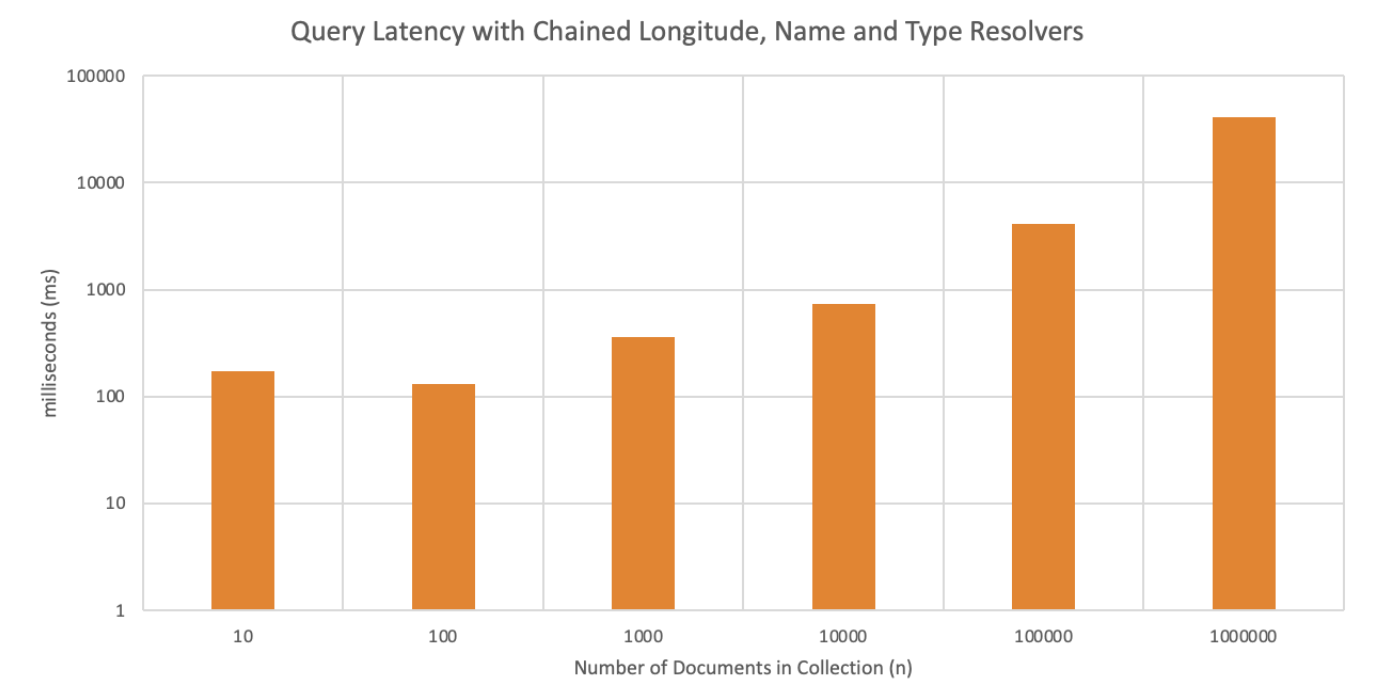}
    \captionsetup{width=.8\linewidth}
    % \decoRule
    \caption{GraphQL query with name=Howard, type=level2, 47.5 $\le$ latitude $\le$ 48.0, -122.5 $\le$ longitude $\le$ -122.1.  Time to complete all the operations shown in milliseconds (ms) across all 6 collections.}
    \label{fig:gql-query}   
\end{figure}

In order to characterize the effect that GraphQL middleware has on latency, a GraphQL server defines an API to reproduce the Firestore query and reduction in Figure \ref{fig:firestore-read-lat}.  The client sends the GraphQL query: 

\begin{lstlisting}[basicstyle=\ttfamily\small]
    query {
        latitude(
            num: 1000000,
            latmin: 47.5,
            latmax: 48.0,
            longmin: -122.5,
            longmax: -122.1,
            name: 'Howard',
            type: 'level2') 
        {
            longitude {
                name {
                    type {
                        id
                        name
                        address
                        latitude
                        longitude
                        type
                    }
                }
            } 
        }  
    }
\end{lstlisting}

\noindent to the GraphQL API and converts to a tree syntax as shown in Figure \ref{fig:gql-tree-resolver}.  The top level field is \emph{ latitude} and triggers a resolver with the Firestore database query.  The results populate the data tree, which is then traversed with each child field (\emph{longitude}, \emph{name} and \emph{type}) resolver that further filters based on the field query conditions. The time to resolve the queries for all six collections is shown in Figure \ref{fig:gql-query}.  The time complexity for this query is near-linear, although there is a drop in latency at $n = 100$.  The drop occurs since the connection setup time contributes more to the latency in the $n=10$ case.

%% file: Chapters/Chapter5.tex
\section{Discussion and Future Work}\label{sec:discussion-future}
In Section \ref{sec:introduction}, we state that the evaluation of the persistent storage solution would be based on:
\begin{enumerate}
    \item Latency performance.
    \item Cost.
    \item Flexibility and scalability.
\end{enumerate}
We discuss these criteria for both the Firebase and MySQL solutions in the following sections.

\subsection{Latency}\label{sub:latency}
In this section, we show an analysis to estimate latency based on the size of the data set $n$ and the size of the response set $r$.  The analysis is performed on experimental latency results from both the Firestore and MySQL databases with the aim of providing a predictive model for each.  The models estimate latency in milliseconds.

The goal of the model is to find a relationship between three variables: $l$ (latency), $n$ (data size), and $r$ (response size).  The dependent variable $l$ is related to the independent variables $n$ and $r$.  A Multiple Linear Regression could model this relationship accurately if the predictors are uncorrelated.  However, Table \ref{tab:mock-data} shows that the mock data set is based on a small pattern that repeats randomly.  As such, we expect a high correlation between $n$ and $r$ (data-based multicollinearity) and may need to focus only on a single predictor.  The Multiple Linear Regression takes the general form of:

\begin{equation}\label{eq:multi-linear}
    Y = \beta_0 + \beta_1 X_1 + \beta_2 X_2
\end{equation}

$Y$ is the output or dependent variable.  The $X$ terms represent the corresponding input or independent variables. Each predictor (input) has a corresponding slope coefficient $\beta$.  The first $\beta$ ($\beta_0$) represents the intercept constant.  In our case, we can restate this equation more specifically as Equation \ref{eq:multi-linear2}.  The determiner method \cite{Hand2018,Moon2019,Eyada2020} is used to solve this equation as shown in Equation \ref{eq:determiner}.

\begin{equation}\label{eq:multi-linear2}
    l = \beta_0 + \beta_1 n + \beta_2 r
\end{equation}

\begin{equation}\label{eq:determiner}
\begin{vmatrix}
    l & n & r & 1 \\
    \sum_{i=1}^{n}l & \sum_{i=1}^{n}n & \sum_{i=1}^{n}r & n \\
    \sum_{i=1}^{n}nl & \sum_{i=1}^{n}n^2 & \sum_{i=1}^{n}nr & \sum_{i=1}^{n}l \\
    \sum_{i=1}^{n}rl & \sum_{i=1}^{n}nr & \sum_{i=1}^{n}r^2 & \sum_{i=1}^{n}r
    \end{vmatrix}
    = 0
\end{equation}\\

Rather than solving Equation \ref{eq:determiner} by hand, we use the Microsoft Excel Data Analysis tool \cite{Excel} to calculate the $\beta$ coefficients on the following data. 

\subsubsection{Firestore Analysis}\label{sub:firestore-analysis}
We will focus only on the read operation utilizing the \emph{ latitude} range condition shown in Table \ref{tab:firestore-regression}.  The other fields such as \emph{longitude}, \emph{name} and \emph{type} must be filtered on the client side and are included in the latency values.  The read and filter is the most applicable scenario in the energy marketplace.  Create, update, and delete operations occur less frequently and are not as relevant to the marketplace performance and user experience.

\begin{table}[H]
    \captionsetup{width=.8\linewidth}
    \caption{Firestore latency for read operations utilizing \emph{latitude} range condition from Figure \ref{fig:firestore-read-lat}.  Both $n$ and $r$ are predictor variables and $l$ is in milliseconds.}
    \label{tab:firestore-regression}
    \centering
    \begin{tabular}{r | r r}
    \toprule
    \tabhead{Latency ($l$)} & \tabhead{Documents ($n$)} & \tabhead{Results ($r$)} \\
    \midrule
    171 & 10 & 3\\
    182 & 100 & 20\\
    234 & 1000 & 181\\
    855 & 10000 & 1970\\
    7718 & 100000 & 20028\\
    66763 & 1000000 & 199958\\
    \bottomrule
    \end{tabular}
\end{table}

We performed regression analysis on the Table \ref{tab:firestore-regression} data in three passes: using both $n,r$ as predictor variables, $n$ only, and $r$ only.  Table \ref{tab:firestore-calculations} summarizes the results and shows the $\beta_0,\beta_1,\beta_2$ coefficients for each pass as well as the standard error (SE) on the predictor coefficients $\beta_1,\beta_2$.

\begin{table}[H]
    \captionsetup{width=.8\linewidth}
    \caption{Summary of regression analysis based on $n$ as a predictor, $r$ and $n,r$ together.  The coefficient values of the model are shown, as well as the standard error for the $n$ and $r$ coefficients.}
    \label{tab:firestore-calculations}
    \centering
    \begin{tabular}{l | l l l l l}
    \toprule
    \tabhead{Variables} & \tabhead{$\beta_0$} &
    \tabhead{$\beta_1$} & \tabhead{SE($\beta_1$)} &
    \tabhead{$\beta_2$} & \tabhead{SE($\beta_2$)} \\
    \midrule
    $n$ & 340.139 & 0.066 & 0.0004 & -- & --\\
    $r$ & 341.394 & -- & -- & 0.333 & 0.002\\
    $n,r$ & 394.892 & -2.761 & 1.118 & 14.141 & 5.593\\
    \bottomrule
    \end{tabular}
\end{table}

Using both $n,r$ in the regression results in a negative $n$ coefficient and a standard error of 5.59, both likely due to the correlation between $n$ and $r$.  With only $n$ as the predictor, the standard error is a low 0.0004.  Using only $r$ as the predictor results in a slightly higher standard error of 0.002.  The differences in slope, $\beta_1,\beta_2$, between the dual predictor case and the single predictors further confirm the high correlation between $n$ and $r$.  Given that a model with a single predictor $n$ yields the smallest standard error, the most accurate predictive model can be represented as:

\begin{equation}\label{eq:firestore-model}
    l = 340.139 + 0.066 n
\end{equation}

\subsubsection{MySQL Analysis}\label{sub:mysql-analysis}
Now we do the same for MySQL read operations querying a specific combination of conditions.  In this scenario, all four conditions are added to the query such that only matching records are returned.  This contrasts with the Firestore results from Section \ref{sub:firestore-analysis} above, where the query engine could only support a single field range condition.  

\begin{table}[H]
    \captionsetup{width=.8\linewidth}
    \caption{MySQL read operation latency while querying \emph{latitude, longitude} range and \emph{name, type} equality condition from Figure \ref{fig:mysql-read-conditions}.  Both $n$ and $r$ are predictor variables and $l$ is in milliseconds.}
    \label{tab:mysql-regression}
    \centering
    \begin{tabular}{r | r r }
    \toprule
    \tabhead{Latency ($l$)} & \tabhead{Records ($n$)} & \tabhead{Results ($r$)} \\
    \midrule
    101 & 10 & 1\\
    80 & 100 & 0\\
    84 & 1000 & 2\\
    122 & 10000 & 32\\
    156 & 100000 & 403\\
    905 & 1000000 & 3874\\
    \bottomrule
    \end{tabular}
\end{table}

Performing the same analysis on Table \ref{tab:mysql-regression} results in the coefficients and standard errors shown in Table \ref{tab:mysql-calculations}.

\begin{table}[H]
    \captionsetup{width=.8\linewidth}
    \caption{Summary of regression analysis based on $n$ as a predictor, $r$ and $n,r$ together.  The coefficient values of the model are shown, as well as the standard error for the $n$ and $r$ coefficients.}
    \label{tab:mysql-calculations}
    \centering
    \begin{tabular}{l | l l l l l}
    \toprule
    \tabhead{Variables} & \tabhead{$\beta_0$} &
    \tabhead{$\beta_1$} & \tabhead{SE($\beta_1$)} &
    \tabhead{$\beta_2$} & \tabhead{SE($\beta_2$)} \\
    \midrule
    $n$ & 90.816 & 0.0008 & 0.00002 & -- & --\\
    $r$ & 90.58 & -- & -- & 0.21 & 0.005\\
    $n,r$ & 92.457 & 0.006 & 0.003 & -1.33 & 0.818\\
    \bottomrule
    \end{tabular}
\end{table}

Again, as we saw in the Firestore case, the correlation between $n$ and $r$ resulted in both a negative slope coefficient on one predictor and a higher standard error when calculating a regression with both predictors.  With a single predictor regression, $r$ has a standard error of 0.005 while $n$ is smaller at 0.00002. Thus, the three-pass regression leads to the most accurate predictive model of:

\begin{equation}\label{eq:mysql-model}
    l = 90.816 + 0.0008 n
\end{equation}

\subsubsection{GraphQL Analysis}\label{sub:graphql-analysis}
For the GraphQL analysis, we will introduce a new predictor variable $r'$ that represents the set of matched results after filtering through the tree.  For this query, the resolvers for each of the four fields are chained.  The root field, \emph{latitude}, sends a request to the Firestore API and populates the data tree with $r$ charger nodes, as shown in Table \ref{tab:gql-regression}.  Each of the next three resolvers (\emph{longitude}, \emph{name} and \emph{type}) then traverse the tree and only return those nodes matching the conditions.  The final set of nodes returned after the traversal is $r'$.  

\begin{table}[h]
    \captionsetup{width=.8\linewidth}
    \caption{GraphQL latency for read operations utilizing \emph{latitude} range condition from Figure \ref{fig:gql-query}.  Predictor variables are $n$, $r$ and $r'$ while $l$ is in milliseconds.}
    \label{tab:gql-regression}
    \centering
    \begin{tabular}{ R{0.1\linewidth} | R{0.2\linewidth} R{0.1\linewidth} R{0.1\linewidth}}
    \toprule
    \tabhead{Latency ($l$)} & \tabhead{Documents ($n$)} & \tabhead{Latitude Results ($r$)} & \tabhead{Matched Results ($r'$)} \\
    \midrule
    174 & 10 & 3 & 1\\
    131 & 100 & 20 & 0\\
    362 & 1000 & 181 & 2\\
    729 & 10000 & 1970 & 32\\
    4150 & 100000 & 20028 & 403\\
    41000 & 1000000 & 199958 & 3874\\
    \bottomrule
    \end{tabular}
\end{table}

\begin{table}[h]
    \captionsetup{width=.8\linewidth}
    \caption{Summary of regression analysis based on $n$ as a predictor, $r$, $r'$ and $n,r,r'$.  The coefficient values of the model are shown, as well as the standard error for the $n$, $r$, and $r'$ coefficients.}
    \label{tab:graphql-calculations}
    \centering
    \begin{tabular}{p{0.07\linewidth} | p{0.07\linewidth}
    p{0.07\linewidth} p{0.07\linewidth} p{0.07\linewidth} p{0.07\linewidth} p{0.07\linewidth} p{0.07\linewidth}}
    \toprule
    \tabhead{Var} & \tabhead{$\beta_0$} &
    \tabhead{$\beta_1$} & \tabhead{SE($\beta_1$)} &
    \tabhead{$\beta_2$} & \tabhead{SE($\beta_2$)} &
    \tabhead{$\beta_3$} & \tabhead{SE($\beta_3$)} \\
    \midrule
    $n$ & 205.35 & 0.04 & 0.0001 & -- & -- & -- & --\\
    $r$ & 206.15 & -- & -- & 0.20 & 0.0007 & -- & --\\
    $r'$ & 192.67 & -- & -- & -- & -- & 10.53 & 0.055\\
    $n,r,r'$ & 147.00 & 1.98 & 0.34 & -10.0 & 1.82 & 15.80 & 5.05 \\
    \bottomrule
    \end{tabular}
\end{table}

After performing a four-pass regression (we now include $r'$ as a predictor), the coefficients and standard errors are shown in Table \ref{tab:graphql-calculations}.  The results are similar to those of both Firestore and MySQL in that using $n$ as a single predictor yields the lowest standard error while combining the predictors results in a negative coefficient as well as higher standard errors due to the correlation between the predictors.  Thus, our most accurate predictive model is as follows:

\begin{equation}\label{eq:graphql-model}
    l = 205.35 + 0.04 n
\end{equation}

\subsection{Cost}\label{sub:cost}
To objectively compare the cost of both solutions, we focus on the worst-case scenario storing $10^6$ charger data objects over one month.  We calculate the cost to create, store, and read the entire dataset.  Each charger object create and read is performed in an individual operation.  The average raw data object size is 76 bytes, giving a stored data size of 76 MB.  Here is the breakdown:

\subsubsection{Firestore}\label{sub:firestore-cost}
Firestore only bills per use, that is, per CRUD operation and per unit of storage \cite{FirestorePricing}.  Note that the experiment is performed in a single day, yet the service remains up for one month.  Although the raw data added to the $10^6$ collection is 76 MB, the Firestore stored data quota shows that 1.9 GB were used.  The indexes for each field and the auxiliary metadata consume the extra storage for the collection.
\begin{table}[H]
    \captionsetup{width=.8\linewidth}
    \caption{Firestore cost estimate to create, store and read $10^6$ documents (76 MB) in one month. The second column is the free quota per day, while the third is the price per unit thereafter.  The fourth column is the amount billed after the free tier, while the fifth column represents the total cost.}
    \label{tab:firestore-cost}
    \centering
    \begin{tabular}{l l l l l}
    \toprule
    \tabhead{} & \tabhead{Free} & \tabhead{Price} & \tabhead{Used} & \tabhead{Total} \\
    \midrule
    Document writes & 20,000 & \$0.09/100,000 & 980,000 & \$0.882 \\
    Document reads & 50,000 & \$0.03/100,000 & 950,000 & \$0.285 \\
    Document storage & 1 GB & \$0.15/GB/month & 0.9 GB & \$0.135 \\
    Internet ingress & all & 0 & 0 & \$0.00 \\
    Internet egress & 10 GB & \$0.12/GB & 0 & \$0.00 \\
    Total & & & & \textbf{\$1.302} \\
    \bottomrule\\
    \end{tabular}
\end{table}

As shown in Table \ref{tab:firestore-cost}, the total cost of a one-month experiment is \$1.30.  Note that this cost is only incurred during use. CRUD operations and data storage are only billed after the free daily quota is exceeded.

\subsubsection{MySQL}\label{sub:mysql-cost}
Cloud SQL MySQL follows a different billing model. Costs are incurred for the dedicated compute in conjunction with the corresponding storage \cite{MySQLPricing}.  In our experiments, the service is provisioned with 1 vCPU, 614.4 MB memory, and 10 GB SSD storage for one month.

\begin{table}[H]
    \captionsetup{width=.8\linewidth}
    \caption{MySQL cost estimate to create, store and read $10^6$ documents (76 MB) for one month.  The majority of the billing (excluding network egress) is per resource, not per transaction.}
    \label{tab:mysql-cost}
    \centering
    \begin{tabular}{l l l l l}
    \toprule
    \tabhead{} & \tabhead{Free} & \tabhead{Price} & \tabhead{Used} & \tabhead{Total} \\
    \midrule
    vCPU & 0 & \$30.149/unit & 1 & \$30.15 \\
    Memory & 0 & \$5.11/GB & 0.614 & \$3.14 \\
    Storage & 0 & \$0.17/GB & 10 & \$1.70 \\
    Internet ingress & all & 0 & 0 & \$0.00 \\
    Internet egress & 0 & \$0.19/GB & 0.076 & \$0.01 \\
    Total & & & & \textbf{\$35.00} \\
    \bottomrule\\
    \end{tabular}
\end{table}

\begin{figure}[H]
    \centering
    \includegraphics[width=1.0\linewidth, keepaspectratio]{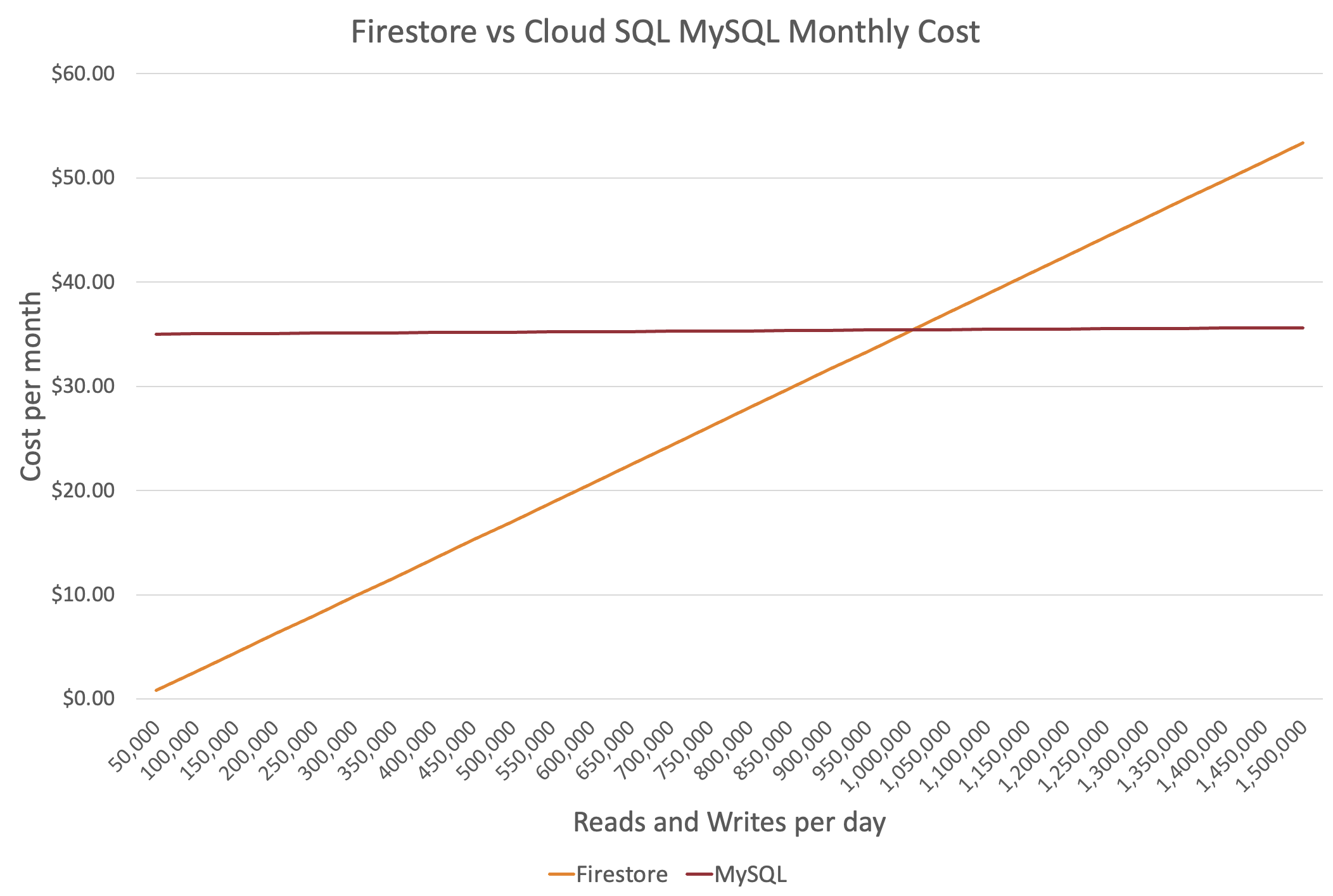}
    \captionsetup{width=.8\linewidth}
    % \decoRule
    \caption{A monthly cost comparison between Firestore and Cloud SQL MySQL.  The x-axis represents the number of read and write transactions (each) per day.  Only the data transferred is factored into the cost.  Other storage costs for index and metadata are ignored in this comparison.}
    \label{fig:cost-comparison}   
\end{figure}

Table \ref{tab:mysql-cost} shows that our costs to run this service for one month are \$35.00.  As we are provisioning a perpetual server (not serverless), costs are incurred regardless if the service is used or not.  The $10^6$-record table represents 76 MB of raw data, but consumes 1.12 GB of database storage due to index and metadata overhead.  MySQL's overhead is significantly less than Firestore's 1.9 GB needed to store $10^6$ charger objects.  In our experiments, database access is restricted to a limited scenario, resulting in Firestore costing less than the MySQL solution.  However, in real-world use, it is possible that the number of transactions will increase to a point where a transaction-based pricing model is more costly than a perpetual resource-based one.  Keeping the same experimental setup and increasing both reads and writes to 1,000,000 per day (each) results in a monthly Firestore cost of \$35.20 which, thereafter, exceeds Cloud SQL MySQL.  Figure \ref{fig:cost-comparison} shows the growth of Firestore costs as the number of transactions increases, while MySQL remains more constant. 

The described limited scenario is suitable for a development environment.  The experiments established a billing period of one month with a large number of read and write transactions occurring in a single day.  Outside of these transactions, Firestore does not incur any costs and is thus convenient (and economical) for a start-and-stop style of iterative development.  The cost can also be reduced with Cloud SQL by stopping the compute service while it is not being used for active development.  However, IP address reservations and storage charges still continue even while the compute instance is paused.  In a production environment, a \$35/month overhead to run a MySQL server is nothing.  Schema-enforced typing may also reduce developer costs by reducing the need for type checking in the application code. 

\subsection{Flexibility and Scalability}\label{sub:flexibility}
Analyzing flexibility and scalability is a more subjective discussion (without empirical data) based on the system architecture as detailed in Chapter \ref{sec:architecture}.  To recap, the NoSQL, non-relational, document-model implemented by Firestore brings the following benefits:
\begin{enumerate}
    \item Stored documents are schemaless, allowing new fields and a changing data shape as development of the marketplace progresses.
    \item Data is stored as a document object with a key-value.  Each document has no relation defined to any other.  Thus, horizontal scaling can be easily performed by distributing documents across multiple nodes, zones, and regions.  This can be done as the database and client demands grow without requiring any service outage.
    \item Separating the management layer (DBMS) from the storage engine allows for easy implementation of a serverless offering.  No dedicated perpetual server is needed.  Firestore listens for client requests on a behind-the-scenes shared compute cluster and only charges per transaction.
\end{enumerate}    

In contrast, Cloud SQL MySQL deploys a server-based instance that is always consuming compute resources along with associated billing.  Tables must define a schema, and any updates to this schema require a migration of the entire table or the creation of a new table.  The migration case will typically involve service downtime as well as developer effort.  Finally, as the table grows in data size and number of records, distribution across multiple nodes requires a complex sharding process that involves splitting the table into child objects and recreating all affected relations.  This sharding will also typically involve downtime for users.  

\subsection{Discussion}\label{sub:discussion}
In this section, we summarize the significant results and compare.  In Sections \ref{sub:exp-non-rel} and \ref{sub:exp-rel}, we deliberately chose to ignore delete operations and focused on create, update, and read.  The reads were divided into single operations querying through a primary key and complex queries involving \emph{latitude}, \emph{longitude}, \emph{name} and \emph{type} conditions.  In the Firestore case, only one of these conditions may be an inequality (range).  MySQL is able to process queries that contain multiple range conditions.  This restriction means that Firestore is forced to return a larger response set than needed and perform further data reduction on the client.

The create, update, and delete operation performance is not as relevant to the energy marketplace.  With only a small group of participants (probably one) performing any of these operations at a time, it is unlikely that a small increase in latency would affect the user experience.  Atomicity, consistency, and isolation are more critical due to potentially overlapping updates.  However, read performance is very relevant.  Querying small datasets $n$ with small response sets $r$ is fast regardless of the database used.  In the larger datasets, we begin to see the differences between the NoSQL and SQL solutions.

To illustrate, take the use case of 1,000,000 chargers.  The Firestore solution stores each as a document, while MySQL is a record within a table.  MySQL is able to query these 1 million records for our specific combination of conditions (two inequality and two equality) and return a matched 3,874 records in $\le$ 1 second (see Table \ref{tab:mysql-regression}).  Firestore queried the same data distributed as 1 million documents in a collection using  only the \emph{latitude} condition.  It took 66 seconds to respond with 199,958 documents, then further reduce to match the \emph{longitude, name, and type} conditions to get the desired 3,874 hits.  MySQL, with its more sophisticated SQL query engine, clearly has an advantage here.      

Table \ref{tab:firestore-read-metrics} shows that, in our experimental setup, Firestore read operations were able to average 19 operations per second while transferring 360 kilobytes per second.  Table \ref{tab:mysql-read-metrics} shows MySQL achieving 54 operations per second with 114,672 kilobytes per second respectively. Clearly, MySQL has the advantage in read performance, while being averaged across all dataset sizes.  Comparing write operations (create or update) from Table \ref{tab:firestore-write-metrics} and \ref{tab:mysql-write-metrics}, we see that Firestore is able to sustain an average of 25 operations per second while MySQL can do 45 operations per second.  Again, the advantage to MySQL.  Additionally, the size of the table consumes server resources and has a finite limit.  The storage ceiling may result in degraded performance at the upper limit that requires hardware upgrades or sharding to other nodes.  Even with the raw latency advantage demonstrated by MySQL in our experiments, Firestore is preferable for expanding the marketplace long-term, as it is the more consistent solution without any known upper limit.

For the purpose of these experiments, all data associated with a charger is stored in a single document or record.  As the marketplace expands, additional data fields, such as user details, may be required.  In Firestore, creating a new document in a "Users" collection to store a picture, charging history, and user details makes the most sense.  There is a feature in Firestore that allows the "Users" to be a subcollection of the chargers collection.  As such, the charger document may contain a reference to the appropriate "Users" document.  All CRUD operations would need to be duplicated to access both documents.  The latency numbers would need to increase accordingly.  If a subcollection is not possible, another strategy is to duplicate the data in both documents.  Again, CRUD operations must be duplicated and persistent storage is increased. 
 MySQL handles this scenario much better through its relations.  The "Users" data can be located in another table and linked to the charger record through a relationship.  The SQL engine dereferences the relation and automatically performs a JOIN for queries that target the charger record.  The latency, storage, and development complexity are minimized compared to Firestore.

Cost is a huge factor as described in Section \ref{sub:cost}.  Our limited, month-long scenario that created, stored, and read one million data objects resulted in a Firestore estimated cost of \$1.30 while MySQL costs \$35.00.  The NoSQL document model database lends itself to serverless operation and shows a significant advantage in providing a pay-as-you-use service.

The insertion of GraphQL middleware provides a significant simplification of the client code.  A very simple query string is defined that represents the desired response data structure.  This query is sent as a single request to the GraphQL API, which filters the response data to an exact match of query.  Filtering eliminates the need for data reduction on the client side to sort through the response data.  Additionally, all fields, field arguments, and response data are strongly typed providing enhanced safety.  These types may evolve over time and can be handled through versioning of the resolver functions.  GraphQL does not provide much latency improvement, as the bottleneck step of querying a single range field within Firestore still exists.

\subsection{Future Work}\label{sub:future-work}
In Section \ref{sub:discussion}, we discussed how Firestore has an advantage with cost and scalability, yet it is at a disadvantage with query latency as result sets increase in size.  Placing a high-performance GraphQL middleware layer provides some optimization, yet does not solve the bottleneck of Firestore's inability to query multiple range fields simultaneously.

A strategy for latency optimization is to organize the collections in such a way as to reduce the size of the result set.  Geohashing \cite{GeoHashing} provides a framework to numerically combine latitude and longitude values.  A single hash value can be queried from the database within a range that represents the desired geographical radius.      

%% file: Chapters/Chapter6.tex
\section{Conclusion}\label{sec:conclusion}
The peer-to-peer energy marketplace for EV charging is a hybrid of Edge devices and Cloud services.  Within these services, persistent data storage is critical.  The details of each charger device must be stored along with user details, session history, billing information, user ratings, amenities, and tourism information.  A Cloud-hosted database can persistently store the data, support performant queries, and provide atomicity, consistency, isolation, and durability (ACID) guarantees along with a convenient pay-as-you go billing model.  The nature of the marketplace workflow causes the create, read, update, and delete (CRUD) operations to not have equal requirements.  For example: create, update, and delete operations are performed by a single user at a time, thus resulting in eventual consistency being acceptable.  Per the consistency, availability, and partition tolerance (CAP) theorem trade-off, the marketplace prioritizes availability and partition tolerance.  However, read operation performance is critical, as it involves multiple users querying the same dataset within a mobile setting.  Additionally, these read operations must be able to support querying multiple range (i.e. inequality condition) fields simultaneously, overlapping in time and competing for memory and CPU resources.    

This study focuses on two popular database services provided by Google to solve the persistent storage research problem for the energy marketplace.  Firestore is a NoSQL, document-model, non-relational and serverless database service while Cloud SQL MySQL is a perpetual server, table-based, relational, and SQL.  Each has its own features and characteristics which we evaluate based on the following criteria:

\begin{enumerate}
    \item Latency performance.
    \item Cost.
    \item Flexibility and scalability.
\end{enumerate}

The SQL database is the established solution, but it has shortcomings in terms of cost and scalability.  The shortcomings lead to our research question: can the Firestore solution be used to support the energy marketplace framework?  We dig into Firestore's architecture and identify these benefits:

\begin{enumerate}
    \item Schemaless data storage provides the flexibility to change document content (data shape) without requiring the complexity and outage to migrate the entire table.  This raises data safety concerns, since a schema is not enforced on incoming data, leading to possible data errors or security vulnerabilities.  In addition, handling changes in the shape of the data may add additional complexity in development.
    \item Each document is stored as a data object independent of the rest.  This granularity without relationship links means that physical storage can easily be spread over multiple nodes, zones, and regions.  Horizontal scaling can take place without interruption.
    \item The document model allows for an easy separation of the database management system (DBMS) and the storage engine with physical storage.  This enables a serverless offering using shared compute and pay-per-transaction pricing.
\end{enumerate}

The more established MySQL architecture exists as a monolithic set of processes that must exist on a single node.  Any distribution requires a complex sharding of the table(s) and an update of the relations.  However, MySQL provides the following benefits:

\begin{enumerate}
    \item The SQL query engine is powerful and can support multiple column range conditions within a single query.  Response records are an exact match, preventing the need to perform data reduction at the client.
    \item Data can be normalized such that only a single instance exists in the database with multiple relations pointing to it.  The normalization allows faster writes since data does not need to be written in multiple places, as well as reducing physical storage size.
\end{enumerate}

The experiments benchmark the latency performance under different load scenarios.  The experiments mimic the energy marketplace workflow from an Edge client.  CRU tests were performed against six data sets that increased from $10^1$ to $10^6$ individual data structures, each representing a charger device.  Update operations using the primary key averaged 160 ms in Firestore but only 92 ms on MySQL.  The read operations with primary key averaged 190 and 92 ms, respectively.  Firestore was only capable of an averaged data rate of 19 operations/s and 360 kB/s, while MySQL could achieve 54 operations/s and 114,672 kB/s.  An additional read scenario ignored the primary key and instead queried with two inequality conditions (\emph{latitude, longitude}) and two equality (\emph{name, type}).  This combination represents a typical workflow in the marketplace.  For up to 10,000 chargers, both databases returned the matched results in less than 1 second.  However, as the size of the data set $n$ increased, Firestore was hampered by the limitation that its query engine could only support a single inequality at a time.  Thus, we must query for \emph{latitude}, return a much larger-than-necessary response set $r$, and do a client-side data reduction.  The worst-case dataset size of 1,000,000 took an incredible 66,763 ms seconds with Firestore while MySQL required only 905 ms.

To optimize the Firestore solution, further experiments were conducted with a GraphQL middleware between the client and the database.  This addition resulted in a significant simplification of the client code.  The client-side query is written as a string containing the desired fields and their corresponding conditions. The query is written to match the desired response data shape.  The need to sort through the response data and perform data reduction at the client is eliminated.  Additionally, the query fields, arguments, and response data are all strongly typed, which enhances code safety.  As the data shape evolves with future marketplace releases, type changes can be handled within the GraphQL API definition.  However, GraphQL does not provide any improvement in query latency, since the bottleneck step of querying a single range field in Firestore from the GraphQL resolver function is still present.   

MySQL is the clear winner in raw latency performance, especially once complex combinations of conditions are used while querying.  However, for a one-month billing cycle that involves the writing, storing and reading of 1,000,000 charger structures, Firestore costs were \$1.30 while MySQL was \$35.00.  The MySQL server was perpetually deployed and incurred costs regardless of whether there was any user traffic.  With the energy marketplace in its early development stages, the cost difference is a deciding factor.  Additionally, the flexibility of the schemaless document model allows features to be deployed in stages with the persistent data adapting to match.  Finally, as the marketplace grows (we hope!), the serverless document-model seamlessly scales horizontally without limit to accommodate both capacity increase and geographical diversity.